\documentclass[aps,physrev,twocolumn,superscriptaddress]{revtex4-2}
\usepackage{amsmath, amssymb}
\usepackage{bbold}
\usepackage{graphicx}
\usepackage[colorlinks=true,linkcolor=blue,urlcolor=black,citecolor=blue]{hyperref}

\newcommand{\re}{\mathrm{Re}}
\newcommand{\im}{\mathrm{Im}}

\newcommand{\vect}[1]{\vec{#1}}

\begin{document}

\title{Optimal entanglement generation in optomechanical systems via Krotov control of covariance matrix dynamics}

\author{Pengju Chen}
\email{pchen9@stevens.edu}
\thanks{equal contribution}
\affiliation{Center for Quantum Science and Engineering and Department of Physics, Stevens Institute of Technology, Hoboken, New Jersey 07030, USA}

\author{Da-Wei Luo}
\email{dluo2@stevens.edu}
\thanks{equal contribution}
\affiliation{Center for Quantum Science and Engineering and Department of Physics, Stevens Institute of Technology, Hoboken, New Jersey 07030, USA}

\author{Ting Yu}
\email{ting.yu@stevens.edu}
\affiliation{Center for Quantum Science and Engineering and Department of Physics, Stevens Institute of Technology, Hoboken, New Jersey 07030, USA}

\date{\today}

\begin{abstract}

We investigate the optimal control of a continuous variable (CV) system, focusing on entanglement generation in an optomechanical system without utilizing Fock basis cutoffs. Using the Krotov algorithm to optimize the dynamics of the covariance matrix, we illustrate how to design a control objective function to manipulate the dynamics of the system to generate a desirable target state. We showed that entanglement between the macroscopic mechanical mirror and the quantum optical cavity can be reliably generated through imposing the control on the detuning of the external laser field. It has been shown that the control may still be achieved when imposing spectral constraints on the external field to restrict it to low-frequency components. In addition, we systematically study the effects of quantum control on non-Markovian open system dynamics. We observed that memory effects can play a beneficial role in mitigating detrimental impacts of environmental noises. Specifically, the generated entanglement shows a reduced decay rate in the presence of these memory effects.

\end{abstract}

\maketitle

\section{Introduction}

In recent years, the interaction between quantum macroscopic objects and microscopic systems has been a research area of significant theoretical intrigue as well experimental and practical importance~\cite{Frowis2018a, Leggett1980a, Farrow2015a, Caldeira2014a}. For example, the interaction between macroscopic spin systems in the form of magnon modes~\cite{Lachance-Quirion2017a,Li2018a,you_nat1,Zhang2016a,Xu2023a,Luo2021a,*Luo2024a} with various quantum systems have been under extensive study. Another widely successful platform for such studies is the optomechanical systems~\cite{cavoptmbook2014s, Aspelmeyer2014a, Barzanjeh2022a, Meystre2013a,Milburn2011a}, consisting of a mechanical oscillator mode interacting with a quantum cavity mode. Optomechanical system has demonstrated a very versatile experimental realizability~\cite{Rashid2016a, Hoang2016a, Dobrindt2008a} with a wide variety of interesting properties, with applications in quantum information processing~\cite{Stannigel2010a, Fiore2011a, Stannigel2012a, Stannigel2011a, yang2015noise,jiang2017chaos}, quantum entanglement generation~\cite{mu2016memory, Liao2014a, Wang2016a, Vitali2007a, Mancini2002a}, and quantum sensing~\cite{Tittonen1999a, Xuereb2011a, Barzanjeh2015microwave}. In such studies, manipulating the system dynamics is often necessary in order to realize a desirable state. Especially, many quantum information tasks~\cite{Nielsen2000a} would require entanglement as a key resource. In the case of continuous-variable quantum information processes, realizing desirable Gaussian states and gates~\cite{Weedbrook2012a,Braunstein2005a}   is an important area of research. In addition, entanglement with macroscopic objects can also have applications in quantum metrology, such as a direct read-out of mechanical motion~\cite{Kotler2021direct} in coupled oscillators, and quantum-enhanced weak force sensing~\cite{Cosco2021a,Weiss2021a}.

To date, there is a wide range of quantum control protocols available~\cite{cong2014control,glaser2015training,koch2016controlling,borzi2017formulation,d2021introduction}, such as stimulated Raman adiabatic passage (STIRAP)~\cite{Vitanov2017a}, chopped random basis~\cite{Doria2011a,Muller2022a} and gradient ascent methods~\cite{Khaneja2005a}. Optimal controls using machine learning tools have also been recently proposed~\cite{Goerz2022a,Niu2019a,Sivak2022a}. Among the various control strategies, the Krotov control~\cite{Konnov1999b,Reich2012v,Tannor1992h, Jager2014a,Goerz2019w} is a versatile algorithm that can be applied to Bose-Einstein condensate~\cite{sklarz2002loading}, photochemistry~\cite{Tannor1992h,somloi1993controlled}, non-linear Schr\"{o}dinger equation~\cite{sklarz2002loading, Reich2012v}, and optimal control under open systems~\cite{hwang2012optimal}, as well as controlling towards entanglers~\cite{Watts2015a,Goerz2015a}. For optomechanical system, controls have been proposed using continuous measurement~\cite{Rossi2018a} and feedback control~\cite{miki2023generating}. Entanglement generation using reservoir engineering~\cite{Wang2013a}, pulsed interaction~\cite{Clarke2020a}, via dissipative effects~\cite{Chen2017a} and external pulses~\cite{Kuzyk2013a} have also been reported. However, to implement a numerical implementation for the optimal control of continuous variable (CV) systems~\cite{Weedbrook2012a,Serafini2017a,Agarwal2012a} would generally require some sort of cut-off, such as cut off the Fock state basis larger than some photon number~\cite{Castro2019a,Porotti2022a,Kudra2022a,Cordero2019a}. Such cutoffs are difficult to estimate and may introduce errors due to the imposed limitations. Depending on the initial state and the target state of the control, this approach can also restrict the range of accessible states, confining the study to states with low photon numbers or those residing within a small subspace. In addition, it can also pose issues in many-body systems as the number of particles grows due to the corresponding exponential growth of the Hilbert space.

In this paper we show how to use the Krotov method to control and manipulate a CV system, in focusing on entanglement generation in optomechanical systems.
Specifically, to address the limitations associated with Fock-basis truncation, we propose applying the Krotov control method to the equations of expectation values. We adapt the Krotov control objective function for the equation of motion of the quantum system's covariance matrix, which directly affects the degree of entanglement. This approach allows for the study of quantum system control without the constraints imposed by Fock-basis truncation.

In this paper, we demonstrate the application of the Krotov method to control and manipulate a CV system, focusing specifically on entanglement generation in optomechanical systems. To address the limitations associated with Fock-basis truncation, we propose applying the Krotov control method to the equations of expectation values. By adapting the Krotov control objective function for the equation of motion of the quantum system's covariance matrix, which directly impacts the degree of entanglement, our approach allows for the study of quantum system control without the constraints imposed by Fock-basis truncation. Moreover, we also investigate key properties of the control algorithm, such as convergence behavior and spectral limits, to ensure its efficacy in practical applications. A realistic description of quantum systems necessitates a more general treatment of open systems, where the environmental memory must be considered. We derive the equation of motion governing the covariance matrix in a general non-Markovian open system setting, and examine the robustness of optimal control strategies under these dynamics. The general approach outlined in this paper will be useful for the study of controlling various continuous variable (CV) quantum systems beyond the specific examples discussed. The paper is organized as follows: We first introduce a general outline of how to utilize the Krotov control algorithm on the CV system in Sec.~\ref{sec_ctrlg}, then turn to a more specific example of generating entanglement between the mechanical mode and the quantum cavity in an optomechanical system in Sec.~\ref{sec_optoe}, with a study on the non-Markovian open systems in Sec.~\ref{sec_opens}.

\section{Optimal control of CV systems} \label{sec_ctrlg}

The control of quantum dynamics plays a crucial role in a diverse range of applications such as state transfer and generation~\cite{Rojan2014a,Yu2023a}, realization of quantum gates~\cite{Riaz2019a}, generating quantum entanglement~\cite{Watts2015a,Goerz2015a}, and assisting quantum metrology~\cite{Luo2023a}, to name just a few. Mathematically, the control problem can be stated as fining the control fields $f_i(t)$ such that the time-dependent Hamiltonian
\begin{align}
	H(t) = H_0 + \sum_i f_i(t) H_i \label{eq_ht}
\end{align}
can drive the system dynamics to some target states at a prescribed time $t_f$, where $H_0$ and $H_i$ are time-independent Hamiltonian describing the uncontrolled part and components under control, respectively. To facilitate control of the dynamics of quantum systems, many different protocols have been proposed~\cite{cong2014control,glaser2015training,koch2016controlling,borzi2017formulation,d2021introduction,Vitanov2017a,Doria2011a,Muller2022a}. A common problem facing quantum optimal controls is that there would be an interdependence of the control field and the quantum state. Without loss of generality, we consider here the case of a single control target. Among the various control strategies, one outstanding algorithm is the Krotov method~\cite{Konnov1999b,Reich2012v,Tannor1992h, Jager2014a,Goerz2019w}. In the context of continuous variable systems, this technique has recently been utilized to generate single-mode squeezing~\cite{Halaski2024a} and engineer beam splitter interactions~\cite{Basilewitsch2022a}. By clever mathematical construction, the Krotov control method decouples the interdependence of the control field and the quantum state under control, and uses an iterative algorithm to minimize a target function of the form
\begin{align}
	& J\left[|\varphi^{(i)}(t_f) \rangle, \{f_l^{(i)}(t)\}\right] \nonumber \\
	&= J_T(|\varphi^{(i)}(t_f) \rangle) + \sum_l \int dt g(f_l^{(i)}(t)), \label{eq_ctrlj}
\end{align}
where $|\varphi^{(i)}(t) \rangle$ is the wave functions at the $i$-th iteration at $t$, evolving under the controls $f_l^{(i)}$ of the $i$-th iteration, $J_T$ is a final time objective function to minimize and $g$ is a correction term of the running cost of the control fields, usually taking the form of
\begin{align}
	g( f^{(i)}_l(t) ) = \frac{\lambda_{a,l}}{S_l(t)}(\Delta f_l^{(i)}(t))^2,
\end{align}
where $\lambda_{a,l}>0$ is an inverse step-size, $\Delta f_l^{(i)}(t)= f_l^{(i)}(t) - f_l^{(i-1)}(t)$ is the difference of the control function between the current and last iteration, and $S_l(t) \in [0,1]$ is an update shape function, generally taken as the Blackman window function~\cite{Ohtsuki2003a,Goerz2019w}. The control pulse can then be updated iteratively using
\begin{align}
	\Delta f_l^{(i)}(t) = \frac{S_l(t)}{\lambda_{a,l}} \mathrm{Im} \left[\left\langle \chi^{(i-1)}(t)\left|\frac{\partial H^{(i)}}{\partial f^{(i)}_l(t)} \right| \varphi^{(i)}(t)\right\rangle\right], \label{eq_ctrlup}
\end{align}
where $| \chi^{(i)}(t) \rangle $ is a co-state that evolves `backwards' according to $H^\dagger(t)$, with boundary condition at the final $t_f$ as $|\chi^{(i-1)}(t_f) \rangle = - \partial J_T/\partial \langle \varphi^{(i-1)} (t_f)|$. By construction, the Krotov control ensures the monotonic convergence of the iterative algorithm in that the control objective function Eq.~\eqref{eq_ctrlj} of the current iteration is guaranteed to be smaller than the previous iteration.

In order to drive an initial states to a target state, the target function $J_T$ is usually taken as the infidelity $1-F$ with $F$ being the fidelity between the target state and the evolved state under control. However, with continuous variable systems, one issue is that to facilitate the numerical calculations, the Hamiltonian and the quantum states needs to have some cutoff in the Hilbert space. While in certain cases the cutoff $N_c$ in the Fock space can be estimated, it can still pose issues where such estimation is difficult. One complication lies in that while the cutoff may be estimated for the initial and final states, special care needs to be taken such that this cutoff applies during all times in the evolution, which can be non-trivial to impose. Moreover, especially with many-body systems, if the required cutoff is large, the computation cost can get very expensive, as the Hilbert space grows exponentially with the number of particles $n$ considered as $N_c^n$. This is especially relevant for tasks such as entanglement generation in CV systems, where operators not conserving the total number of excitations such as non-linear interactions are called for~\cite{Pfister2004a,Braunstein2005a,Adesso2007a,He2012a}.

To avoid this issue, we can instead apply the Krotov control to the equations of motion for the expectation values. We consider, for example, a generic quadratic Hamiltonian $H=R_i M_{ij} R_j/2$ (double index implies summation throughout the paper)~\cite{Weedbrook2012a,Plenio2004a} with the canonical position and momentum operators $R = (q_1, q_2,\ldots, q_n, p_1, p_2,\ldots, p_n)$, whose commutation relationship is given by $[R_i, R_j] = i\sigma_{ij}$. With the ordering of canonical operators $R$ chosen here, the $\sigma$ matrix takes the form of
\begin{align}
	\sigma = \begin{bmatrix} 0 & \mathbb{1} \\ -\mathbb{1} & 0\end{bmatrix}.
\end{align}
Here we will be focusing on Gaussian states, whose information can be encapsulated in a covariance matrix~\cite{Braunstein2005a,Agarwal2012a} (CM) $\gamma$, defined through its matrix elements as $\gamma_{ij}=\langle R_i R_j \rangle+ \langle R_jR_i \rangle - 2\langle R_i \rangle \langle R_j \rangle$. The CM evolves according to
\begin{align}
	\partial_t \gamma = \sigma\mathcal{M} \gamma + \gamma [\sigma\mathcal{M}]^T \label{eq_dcm_m}
\end{align}
where the Hamiltonian is symmetrized as $\mathcal{M}=({M+M^T})/{2}$. We can then vectorize the CM by stacking its columns, denoted as $\vect{\gamma}$. Using the Kronecker product property $AOB = [B^T\otimes A] \vect{O}$, the equation of motion Eq.~\eqref{eq_dcm_m} in for the vectorized CM $\vect{\gamma}$ follows
\begin{align}
	\partial_t \vect{\gamma}
	= \left[\mathbb{1} \otimes \sigma \mathcal{M} + \sigma \mathcal{M} \otimes \mathbb{1}\right] \cdot \vect{\gamma}, \label{eq_dcm_v}
\end{align}
which takes the form of a Schr\"odinger-like equation with a non-Hermitian effective Hamiltonian $-i H_{\rm sys} = \left[\mathbb{1} \otimes \sigma \mathcal{M} + \sigma \mathcal{M} \otimes \mathbb{1}\right]$. For the optimal control towards a target CM, we will now use this effective Hamiltonian in the pulse update equation Eq.~\eqref{eq_ctrlup}, where the vectorized CM $\vect{\gamma}$ replaces the wave functions $|\varphi (t)\rangle$.

With the linear equation of motion for the CM in place, we can now construct the target functional for the control as the vector distance for the real-valued CM with the squared 2-norm $d_2(x,y) = (x-y)\cdot(x-y)$,
\begin{align}
	J_T &= d_2(\vect{\gamma}_T, \vect{\gamma}(t_f)) \nonumber \\
		&= ||\vect{\gamma}_T ||^2 + || \vect{\gamma}(t_f) ||^2 - 2 \vect{\gamma}_T \cdot \vect{\gamma}(t_f), \label{eq_jtd2}
\end{align}
where $\gamma_T$ is the target CM, and $\gamma(t_f)$ is the controlled CM at the end of the evolution at time $t_f$. The corresponding boundary condition for the backwards evolving costate $\vect{\chi}$ for the next iteration is given by
\begin{align}
	\vect{\chi}(t_f) = -\frac{\partial J_T}{\partial \vect{\gamma}(t_f)} = 2 \left[ \vect{\gamma}_T - \vect{\gamma}(t_f) \right].
\end{align}
This approach is akin to formulating the control with vectorized density operators evolving under a master equation~\cite{Goerz2019w}, with a control target using the Hilbert-Schmidt distance between the density operators. It is also worth pointing out that in contrast to the density operators, the CM is real-valued and the distance is generally not upper-bounded, which calls for a careful tuning of the control parameters such as the step size $\lambda_{a, l}$ and the update function $S_l(t)$ to ensure a monotonic convergence.

\section{Entanglement generation in an optomechanical system} \label{sec_optoe}

\begin{figure}
    \centering
    \includegraphics[width=.4\textwidth]{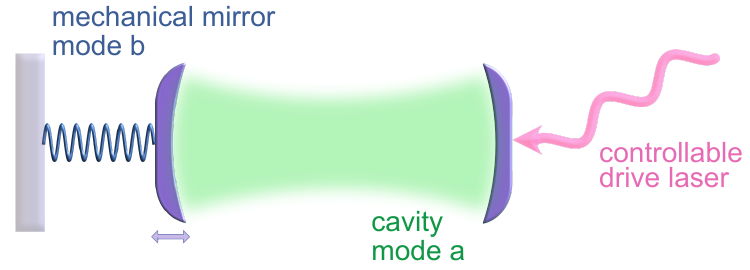}
    \caption{Schematic of the model under consideration. The single-sided optomechanical system consists of a mechanical mirror mode coupled to an optical cavity mode driven by an external coherent laser. The laser is tunable and set to dynamically change the effective detuning $\Delta$ to realize the optimal control.}\label{fig_model}
\end{figure}

As an illustrative example, we consider the problem of entanglement generation in an optomechanical system. The optomechanical system describes an interesting coupling between the quantized cavity mode and a macroscopic movable mirror, which can enable a wide variety of studies on the interaction between microscopic quantum systems and macroscopic objects, such as macroscopic entanglement generation~\cite{Clarke2020a,Wang2016a,Vitali2007a,Wang2013a}. In a typical optomechanical system~\cite{Milburn2011a,Meystre2013a,Aspelmeyer2014a,cavoptmbook2014s,law1994effective,mu2016memory} there exists a parametric coupling between the displacement of the mechanical mode, and the Hamiltonian of the system can be written as (taking $\hbar=1$)
\begin{equation}
	H = \omega_c \tilde{a}^\dagger \tilde{a} + \omega_m \tilde{b}^\dagger \tilde{b} + g \tilde{a}^\dagger \tilde{a}(\tilde{b}+\tilde{b}^\dagger)+\Omega_d (\tilde{a} e^{i\omega t}+h.c.),
\end{equation}
where $\tilde{a}^\dagger$ $(\tilde{a})$ and $\tilde{b}^\dagger$ $(\tilde{b})$ are creation (annihilation) operators of the cavity light field and mechanical mode respectively, with frequencies $\omega_c$ and $\omega_m$,
$g$ is the single-photon optomechanical coupling strength, and $\Omega_d$ is the amplitude of the driving laser with frequency $\omega$.
In the strong driving regime, the Hamiltonian $H$ can be linearized as~\cite{Aspelmeyer2014a, cavoptmbook2014s}
\begin{equation}
	H_S =-\Delta a^\dagger a +\omega_m b^\dagger b + G(a^\dagger +a)(b^\dagger +b), \label{eq_hss}
\end{equation}
where $a$ and $b$ are the quantum fluctuations of optical and mechanical modes around their mean values ($\alpha$, $\beta$), with the linearization $\tilde{a} = a +\alpha$ and $\tilde{b} = b +\beta$. $G=\alpha g$ is the effective coupling rate, and $\Delta=\omega-\omega_c+2G^2/\omega_m$ is the modified detuning. The fluctuations can be determined by~\cite{Aspelmeyer2014a, cavoptmbook2014s, mu2016memory} as $\left[i (\omega - \omega_c)-ig(\beta + \beta^*) - \kappa_a\right]\alpha=i \Omega_d$ and $-i \omega_m \beta = ig |\alpha|^2$ with classical cavity leakage rate as $\kappa_a$. This represents possibly one of the simplest models for the optomechanical cavity. To generate entanglement between the mechanical mirror and the optical cavity using the optimal control strategy outlined in the last section, we propose to use a tunable drive laser on the optomechanical system, which may be realized experimentally~\cite{Milne2020a,Bishof2013a}. We show a schematic of the controlled model under consideration in Fig.~\ref{fig_model}. Specifically, by tuning the frequency and amplitude of the external drive laser, one may dynamically change the detuning $\Delta$ in Eq.~\eqref{eq_hss}. Denoting the now time-dependent detuning parameter which would serve as the control function $f(t)$, we have
\begin{align}
	H_c(t) &= f(t) a^\dagger a +\omega_m b^\dagger b + G(a^\dagger +a)(b^\dagger +b) \nonumber \\
		   &\simeq \frac{f(t)}{2} \left[p_c^2 + q_c^2\right] + \frac{\omega_m}{2} \left[p_m^2 + q_m^2\right] + 2G q_c q_m, \label{eq_om_hc}
\end{align}
where we have rewritten the Hamiltonian in the position-momentum basis with $q_c = ( a^\dagger + a )/\sqrt{2}$, $p_c = i ( a^\dagger - a )/\sqrt{2}$ for the cavity mode and similar for the mechanical mode. This is a quadratic Hamiltonian and the $\mathcal{M}$ matrix in the equation of motion for the vectorized CM Eq.~\eqref{eq_dcm_v} can be given by, with $R = \{ q_c, q_m, p_c, p_m \}$, as $\mathcal{M} = \mathcal{M}_0 + f(t) \mathcal{M}_c$ with
\begin{align}
	& \mathcal{M}_0 = \left[\begin{matrix}
			0 & 2G & 0 & 0 \\
			2G & \omega_m & 0 & 0 \\
			0 & 0 & 0 & 0 \\
			0 & 0 &0 & \omega_m
	\end{matrix}\right],
\end{align}
and $\mathcal{M}_c = \mathrm{diag}(1, 0, 1, 0)$.

\begin{figure}
	\centering
	\includegraphics[width=.45\textwidth]{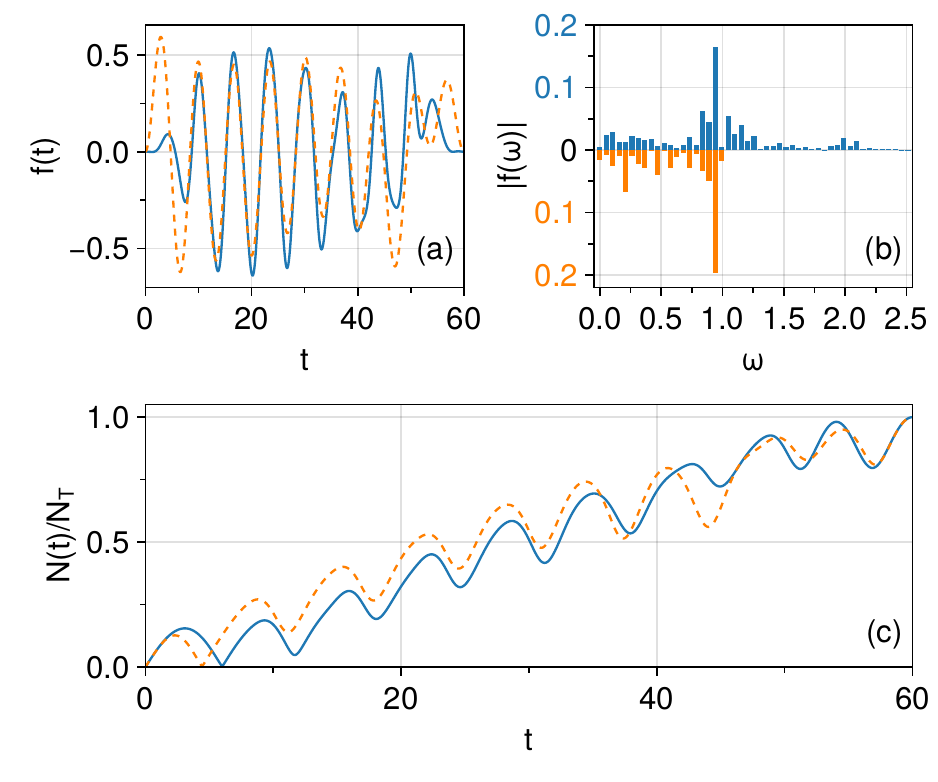}
	\caption{Optimal generation of entanglement in an optomechanical system. Panel (a): the control fields in the time domain, with (orange dashed line) and without (blue solid line) spectral cutoff. Panel (b): The control fields in the frequency domain, showing the absolute values of the amplitudes for control fields with (orange) and without (blue) spectral cutoffs. Panel (c): Dynamics of the log negativity divided by the target negativity, as a function of time.} \label{fig_kcfft}
\end{figure}

Here, the target CM is chosen to be the entangled two-mode squeezed state~\cite{Ekert1989a,Agarwal2012a} given by $S(r) |00 \rangle = \exp\left[r \left(ab-a^\dagger b^\dagger\right)\right] | 00 \rangle$ with a squeezing parameter $r \in R$, where $|00 \rangle$ is the two-mode vacuum state. The explicit expression of the two-mode squeezed state CM can be written in the form of
\begin{align}
	\gamma_T = \left[
		\begin{array}{cccc}
			\cosh(2r) & -\sinh(2r) & 0 & 0 \\
			-\sinh(2r) & \cosh(2r) & 0 & 0 \\
			0 & 0 & \cosh(2r) & \sinh(2r) \\
			0 & 0 & \sinh(2r) & \cosh(2r) \\
		\end{array}
	\right].
\end{align}

The entanglement of the two-mode Gaussian states can be measured by the logarithmic negativity~\cite{Vidal2002a,Horodecki2009a,Simon2000a,Plenio2005a},
\begin{align}
	N = -\sum_i \log_2\left(\min(1, |\lambda_i|)\right)
\end{align}
where $\lambda_i$ are the symplectic eigenvalues (accounting for the 2-fold degeneracy) of the partial-transpose CM $\gamma^{T_B} = P \gamma P$, $P=\mathrm{diag}(1,1,1,-1)$~\cite{Simon2000a,Plenio2004a}. Especially, for the target two-mode squeezed state with $r>0$, we have $N_T=2\log_2(e^r)$.

Choosing system parameters $\omega_m = 1$, $G=0.1\omega_m$, we now use the Krotov control algorithm to generate the control field to drive an initial vacuum state $|00\rangle$ to the target desired two-mode squeezed state with a squeezing parameter $r=1.25$. Experimentally, the single photon coupling $g_0$ has been realized between $10^{-6} \omega_m$ and $10 \omega_m$~\cite{Aspelmeyer2014a}. Thus, our choice of the effective coupling rate $G=0.1 \omega_m$ could be made within experimental viable ranges, where $G=g_0\sqrt{n_c}$ and $n_c$ is the cavity photon number. Notably, coupling rates of similar values has been reported in~\cite{Dare2024a,Groblacher2009a,Rios-Sommer2021a} at $G\sim 0.12 \omega_m$, $G\sim 0.2 \omega_m$ and $G\sim 0.34 \omega_m$. The target negativity $N_T \approx 3.6067$, and the control goal is set to be such that the distance between the evolved CM and the target CM is less than $10^{-4}$, with a control run-time of $t_f = 60$. The initial guess for the control is set to be a constant $f(t) = 0$. The resulting controlled detuning $f(t)$ is shown in Fig.~\ref{fig_kcfft} (a) as the solid blue line. We also show the entanglement dynamics between the mechanical mirror and the optical cavity in Fig.~\ref{fig_kcfft} (c), as the ratio of the negativity of the evolved state divided by the target negativity (solid blue line). It can be seen that the controlled detuning is on the same order-of-magnitude with other parameters of the Hamiltonian, and we are able to reach the target entangled state at the end of the evolution.

\begin{figure}
	\centering
	\includegraphics[width=.45\textwidth]{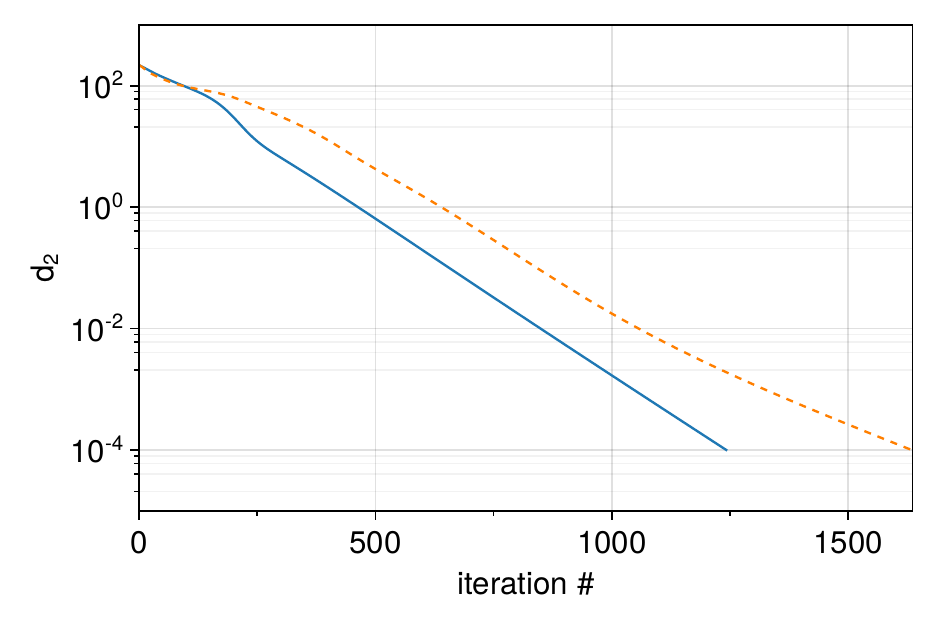}
	\caption{The distance between the final covariance matrix under control and the target covariance matrix, against the Krotov iteration count (log scale). The blue line shows the case where no spectral cutoff is applied, and the orange dashed line is the case where a spectral cutoff is applied after each iteration. It can be seen that while it takes more iterations to reach a target distance, both case are effective control that can roughly approach the target exponentially.} \label{fig_iter}
\end{figure}

We also study the effectiveness of the Krotov iterative algorithm. In Fig.~\ref{fig_iter} we show the squared distance $d_2$ Eq.~\eqref{eq_jtd2} for each iteration. With a search step-size of $1/\lambda_a=1/8000$, we are able to reach the target $J$ in $\sim 1200$ iterations, and we can see that the distance between that evolved state and the target state decreases not only monotonically but also exponentially with the number of iterations, suggesting that the control algorithm is very effective at converging to the desired control function.

\begin{figure}
	\centering
	\includegraphics[width=.3\textwidth]{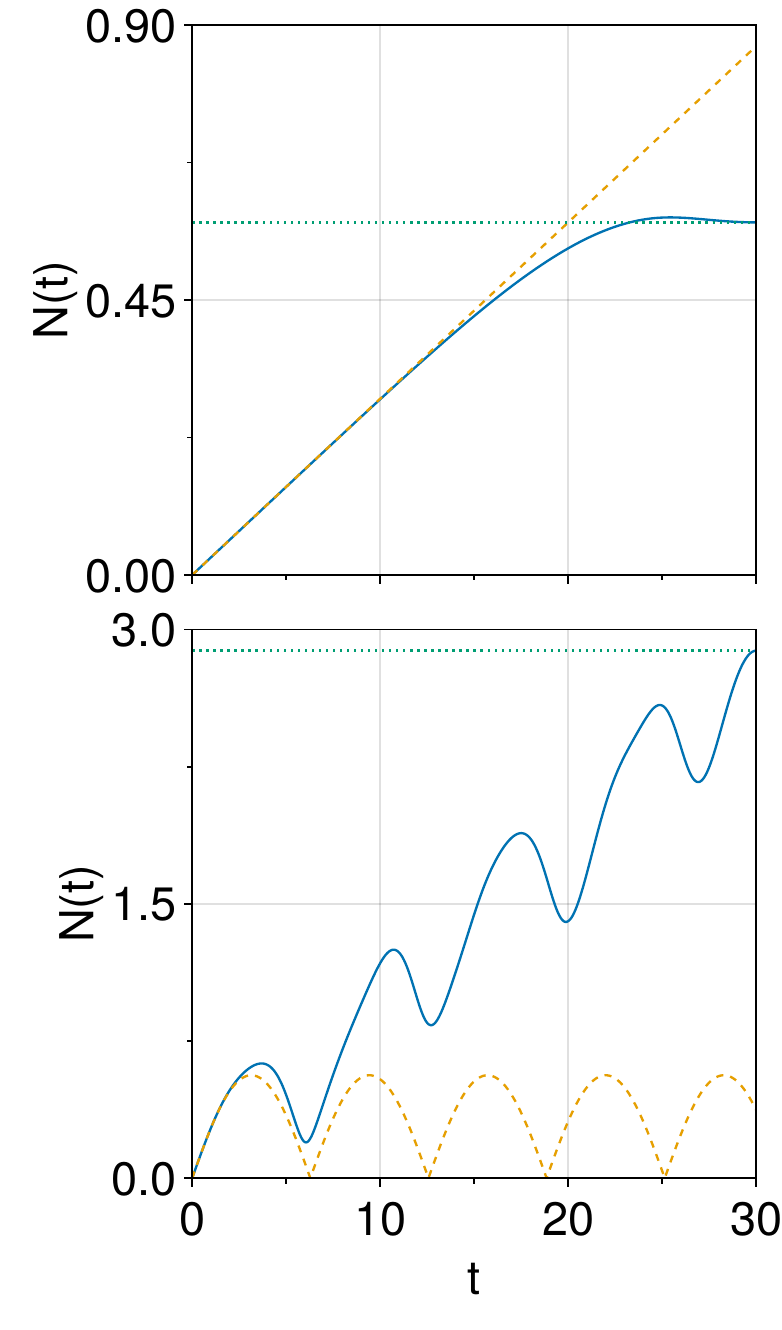}
	\caption{Dynamics in different parameter regions with (blue solid line) and without control (orange dashed line), for a total runtime of $t_f = 30$: upper panel is for the case where $G/\omega_m=0.01 \ll 1$ with a target squeezing factor $r=0.2$ and the lower panel is for $G/\omega_m = 0.1$, with a target squeezing factor $r=1.0$. The green dotted line denotes the target entanglement.}\label{fig-rwax}
\end{figure}

As a study of the robustness of the control, we consider two distinct cases: for weak coupling rates $G \ll \omega_m $, if one lets $f(t) = -\omega_m$ we have an effective Hamiltonian, after rotating out $-\omega_m a ^\dagger a + \omega_m b ^\dagger b$, and applying the rotating wave approximation (RWA), $H' \approx G(ab + h.c.)$ in the interaction picture, equivalent to the two-mode squeezing operator. Going back to the Schr\"odinger picture, the entanglement measured by the logarithm negativity would grow linearly without control (upper panel of Fig.~\ref{fig-rwax}, orange dashed line, with a constant $f(t) = -\omega_m$). Note that in this case, the Krotov control is still valid since we are able to reach the target state at the prescribed runtime $t_f=30$ for a target squeezing factor $r=0.2$ (blue solid line, upper panel of Fig.~\ref{fig-rwax}). On the other hand, when the coupling strength $G$ is larger, the RWA would no longer apply. In such case, the uncontrolled entanglement dynamics may just oscillate while a much higher degree of entanglement can be reached with the Krotov control applied (squeezing factor $r=1.0$, lower panel of Fig.~\ref{fig-rwax}, where for the uncontrolled case we set a constant $f(t) = 0$). It's worth pointing out that for the uncontrolled dynamics in the RWA case, while the entanglement approximately grows linearly in time as the effective squeezing parameter under $H'$ is $Gt$, the linear growth would be limited to a very slow rate since we need $G \ll \omega_m$ for the RWA to hold in this uncontrolled scenario.

\begin{figure}
	\centering
	\includegraphics[width=.42\textwidth]{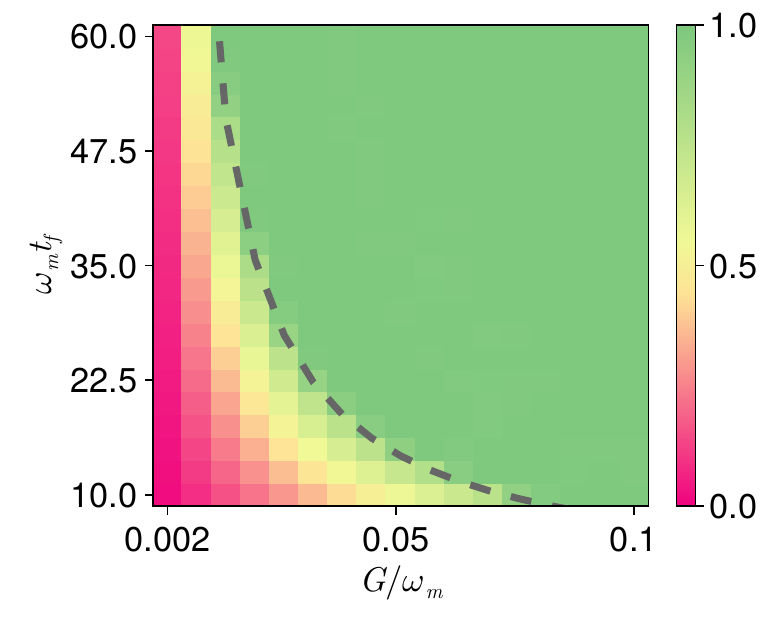}
	\caption{Final relative entanglement $N(t_f)/N_T$ where $N_T$ is the target negativity, as a function of the coupling strength $G$ and runtime $t_f$. The gray dashed line denotes analytical estimation of the quantum speed limit time. The target state is chosen to be a two-mode squeezed state with a squeezing parameter $r = 0.8$.} \label{fig-scangtf}
\end{figure}

In the control of quantum dynamics, one is often interested in how to achieve the control target in the shortest amount of time, a topic generally known as the quantum speed limits (QSL)~\cite{Deffner2017a, Mandelstam1991a, Bhattacharyya1983a}. With its roots in the uncertainty principle, QSL has been rediscovered in various studies such as geometric approaches~\cite{Taddei2013a}. In the context of optimal controls~\cite{Gajdacz2015a}, one may estimate the quantum speed limit time as~\cite{Caneva2009a}
\begin{align}
	T_{\rm QSL} \sim \Delta E_0^{-1} \arccos|\langle \psi(0) | \psi_{T} \rangle|, \label{eq_qsl}
\end{align}
where $|\psi_{T} \rangle$ is the target state and $|\psi(0) \rangle$ is the initial state. For the system under consideration here, the variance $\Delta E_0 = \sqrt{\langle H^2 \rangle - \langle H \rangle^2} = G$ for the initial vacuum state and $H=H_c$ in Eq.~\eqref{eq_om_hc} with an initial guess control $f(t) = 0$ as with the vacuum initial state the term $a ^\dagger a$ has no contribution. The target two-mode squeezed state can be written in the Fock basis as~\cite{Ekert1989a,Schumaker1985a} $S(r)|00 \rangle = (\cosh r)^{-1} \sum_{n=0}^\infty (-\tanh r)^n |n,n \rangle$ so the inner product between the initial and target state is $1/\cosh r$. The QSL time is thus estimated as $T_{\rm QSL} \sim \arccos(1/\cosh r)/G$. In Fig.~\ref{fig-scangtf} we show the relative final entanglement $N(t_f)/N_T$ where $N_T$ is the target negativity, as a function of runtime $t_f$ and the coupling strength $G$, as well as a boundary estimated by the QSL Eq.~\eqref{eq_qsl}. Here, the target state is chosen to be a two-mode squeezed state with a squeezing parameter $r = 0.8$. It can be seen that one can reach the target entangled state whenever the evolution time is larger than the limit imposed by the QSL, while the target state may not be reachable when the evolution time is shorter than the QSL time. In this sense, with the control protocol presented here, we can have an optimal control at the quantum speed limit.

\subsection*{Krotov control with optional spectral cutoff}

In a typical practical scenario, it is often preferable to use a control field that is neither too strong in magnitude nor oscillates too rapidly. While it can be observed clearly in Fig.~\ref{fig_kcfft} (a), we may quantitatively study it by taking a Fourier transformation of the control field. Given that the control field is real-valued, we use the standard discrete cosine transformation (DCT),
\begin{align}
	Y_k = 2 \sum_{j=0}^{n-1} X_j \cos\left[ \pi (j+1/2) k / n \right]
\end{align}
with the inverse
\begin{align}
	W_k &= Y_0 + 2\sum_{j=1}^{n-1} Y_j \cos\left[ \pi j (k+1/2) /n \right] \nonumber \\
		& \equiv f(\omega=0) + 2\sum_{j=1}^{n-1} f(\omega_j) \cos\left[ \omega_j t + \varphi_j \right],
\end{align}
where $\omega_j = \pi j / t_f$, $\varphi_j = {\pi j}/{2n}$, with a logical DFT size $N=2n$ as an overall normalizing factor, and $n=6000$ is the control field's time grid size. In Fig.~\ref{fig_kcfft} (b) we show the absolute values of the control $f$ in the frequency domain.

It can be readily seen that the angular frequencies of the control field is relatively small (here it is smaller than $2$ in our parameter choice), showing vanishingly small amplitudes for larger frequencies: here, the sum of absolute amplitudes for $\omega > 2.56$ is smaller than $\sim 0.00446$. We may also further restrict the spectral limit of the control field. While it may require solving for a complex Fredholm equation~\cite{Reich2013a}, in simple cases it can suffice to cut off the higher frequencies after each iteration~\cite{Goerz2019w}. Here, we keep only the first $20$ frequencies for the DCT, corresponding to a cut-off frequency of $\sim 1$. To keep both ends of the control field constant $f(0) = f(t_f) = 0$, the control field after the frequency cut is then multiplied update window function $S(t)$ in Eq.~\eqref{eq_ctrlup}. In Fig.~\ref{fig_kcfft} (a) and (b) we show the final control field with the spectral constraint in both time and frequency domain, and the resulting entanglement dynamics is shown in Panel (c) in Fig.~\ref{fig_kcfft}, as dashed orange lines. It can be seen that we are still able to reliably generate the desired entangled state with the spectral cutoff for the control field. The distance $d_2$ in each iteration is displayed in Fig.~\ref{fig_iter}. We observed that while it takes more number of iterations to reach the target distance of $10^{-4}$, the distance between the evolved state and the target state still converges mostly exponentially and keeps the monotonic trend of convergence, showing that this strategy is effective in the case where the spectral constraint is being applied.

\section{Open system effects on the entanglement generation} \label{sec_opens}

\begin{figure}
	\centering
	\includegraphics[width=.47\textwidth]{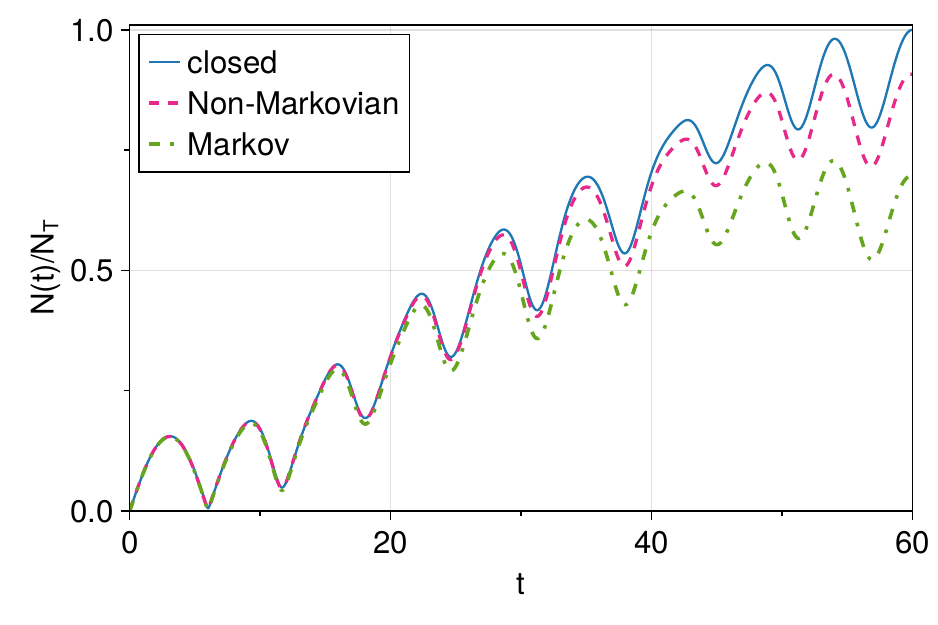}
	\caption{The controlled system entanglement dynamics under open system effects, showing the negativity as a function of time, with system-bath coupling strength $\lambda = 0.1$. The solid blue line represents a reference for the closed system dynamics, while the red dashed line shows the non-Markovian dynamics ($\eta=0.5$), and the green dash-dotted line shows the Markov dynamics case. It can be seen that the entanglement is better preserved when the memory effects of the environment are taken into consideration.}\label{fig_opent}
\end{figure}

In this section, we will study how the optimal control of quantum entanglement is affected by the environment effects. It is well-known that entanglement is a fragile resource, highly susceptible to various environmental effects such as decoherence, dephasing, and entanglement sudden death~\cite{Yu2009a,Yu2006a}. To study how the proposed control strategy is robust against the influences of open system effects, we consider the optomechanical system embedded in a bosonic environment,
\begin{equation}
	H_B=\sum_j\omega_j \bar{b}_j^\dagger \bar{b}_j,
\end{equation}
where $\bar{b}_j^\dagger$ ($\bar{b}_j$) are bosonic creation (annihilation) operators. The system-bath interaction is given by
\begin{equation}
	H_I=\sum_jg_j(L\bar{b}_j^\dagger+L^\dagger \bar{b}_j),
\end{equation}
where $L$ describes the damping of the mechanical oscillator and $g_j$ is the coupling strength between the system and the environment. To obtain the reduced system dynamics, we use the quantum state diffusion (QSD) equation~\cite{diosi1998non,strunz1999open,mu2016memory,Yu1999a}, which projects the bath modes onto the Bargmann coherent states to arrive at a stochastic Schr\"odinger equation for a pure state representing a quantum trajectory,
\begin{equation}
	\partial_t |\psi_t(z^\ast)\rangle
	=\bigg[-iH_s + Lz^*_t - L^\dagger \bar{O}(t,z^*)  \bigg]|\psi_t(z_t^\ast)\rangle,
\end{equation}
where $O(t,s,z^*)\psi_t\equiv\frac{\delta\psi_t}{\delta z_s}$ is an ansatz for the functional derivative with the initial condition $O(t, s=t, z^\ast) = L$ and $\bar{O}(t,z^\ast)\equiv\int^t_0ds\alpha(t,s)O(t,s,z^*)$. When the bath spectrum is of the Lorentzian type, the correlation function of the environment $\alpha(t,s)$ is given by
\begin{equation}
	\alpha(t,s)=\frac{\eta}{2}e^{-(\eta+i\Omega)|t-s|},
\end{equation}
where $1/\eta$ represents the memory time and $\Omega$ signifies a central frequency shift. This choice of the correlation function allows us to study how the system behaves under a non-Markovian bath with a continuously tunable strength of the memory effects, with smaller $\eta$ corresponding to stronger memory effects, whereas $\eta \rightarrow \infty$ would lead to a memoryless (Markov) dynamics. The $O$-operator follows the consistency condition
\[
	\partial_t \frac{\delta}{\delta z_s} | \psi_t(z_t^*) \rangle = \frac{\delta}{\delta z_s} \partial_t | \psi_t(z_t^*) \rangle.
\]
The reduced density operator may be obtained by a stochastic average $\rho = M[| \psi_t(z_t^*) \rangle \langle  \psi_t(z_t) |]$, where $M[\cdot]\equiv\int\frac{dz^2}{\pi}e^{-|z|^2}[\cdot]$ represents the average over the noise (a Gaussian process) $z_t$.

\begin{figure}
	\centering
	\includegraphics[width=.4\textwidth]{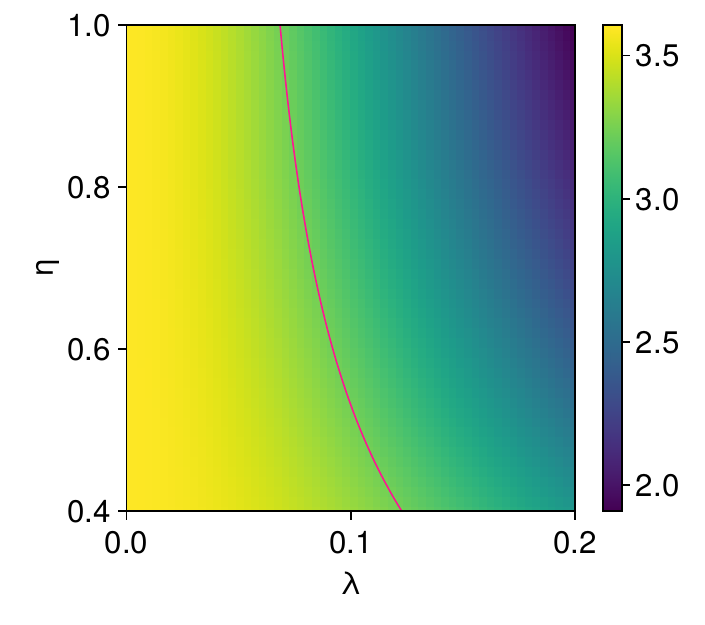}
	\caption{The negativity at $t_f$ as a function of the system-bath coupling strength $\lambda$ and the memory coefficient $\eta$, and the red line shows the boundary where $90 \%$ of the target negativity can be retained.}\label{fig_openhm}
\end{figure}

It can be analytically derived (see Appendix~\ref{sec_appendix_dcm}) that the CM under a general non-Markovian bath of the Lorentzian spectrum may be given by
\begin{align}
	\partial_t \gamma &= \left[\sigma\mathcal{M} + \sigma \Delta \right] \gamma + \gamma \left[\sigma\mathcal{M} + \sigma \Delta\right]^T + 2 \left[\sigma \delta^R \sigma^T \right]
\end{align}
where $\delta_{mn} = l_m^* o_n + o_m^* l_n$, $\delta^R = \re [\delta]$, $\Delta_{mn}=i l_m o_n^* - i l_m^* o_n$, and $L=l_i R_i$, $\bar{O} = o_i R_i$ are the system-bath interaction and the noise-free $\bar{O}$ operator, assuming a linear composition for both operators. In the Markov limit, we have $\bar{O}=L/2$, and the equation above reduces to the well-known result of the CM under Markov baths~\cite{Wiseman2005a,Zhang2017a} with $[\delta^R]_{mn} \rightarrow \re(l_m^* l_n)$, and $\Delta_{mn} \rightarrow \im (l_m^* l_n)$, where they are generally known as the diffusion and drift matrices.

Here, we consider a case where the mechanical mode is coupled to a bosonic bath~\cite{mu2016memory}. To begin with, we assume that only the mechanical mode is dissipative, where system-bath coupling operator is given by $L=\lambda b$, where $\lambda$ denotes the system-bath interaction strength. With $\lambda=0.1$, we plot the open system dynamics of the entanglement under both a non-Markovian environment ($\eta=0.5$) and a Markov environment in Fig.~\ref{fig_opent}. Compared with the ideal closed system dynamics, it can be seen that the entanglement generation can be degraded under decoherence, while the non-Markovian case fares better and can retain a higher degree of entanglement.

\begin{figure}
    \centering
	\includegraphics[width=.4\textwidth]{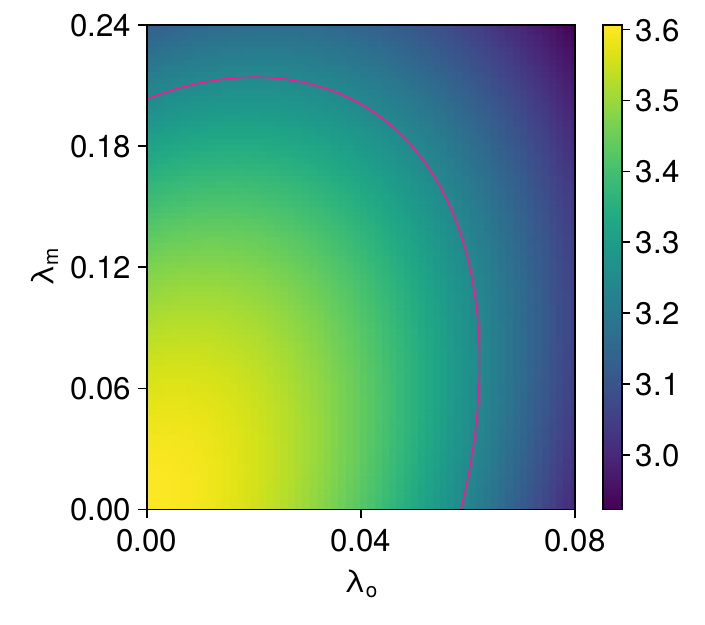}
    \caption{Controlled entanglement at $t_f$ as a function of the system-bath coupling strength of the optical mode $\lambda_o$ and of the mechanical mode $\lambda_m$, taking $\eta =0.2$. Red line shows the boundary where $90 \%$ of the target entanglement is retained.}\label{fig_open_l2}
\end{figure}

To further probe the influences of the environmental noises, we plot the final-time entanglement as a function of the system-bath coupling strength $\lambda$ and the bath memory parameter $\eta$ in Fig.~\ref{fig_openhm}, with a region highlighted where $\sim 90 \%$ of the target entanglement can be retained. It can be seen that a weaker coupling or a stronger memory effects may both help protect the entanglement against the detrimental effects of the noise. The memory effects of non-Markovian environments can also lead to some counterintuitive behaviors. Consider the case where both the optical mode and the mechanical mode are coupled to the bath, with $L = \lambda_o a + \lambda_m b$. The non-Markovian bath would then induce an indirect coupling between the optical and mechanical modes. In Fig.~\ref{fig_open_l2} we show the entanglement at $t_f$ as a function of the system-bath coupling strengths $\lambda_o$ and $\lambda_m$, with $\eta=0.2$. It can be seen that stronger coupling rates to the bath does not necessarily mean the entanglement will decay more in a non-Markovian bath, and in some regions stronger coupling strengths retains the entanglement better, due to the indirect coupling induced by the non-Markovian bath. This feature is generally not observed in Markov baths and highlights how the memory effects of the bath may help in the preservation of quantum entanglement.

\section{Conclusion}

In conclusion, we have proposed an optimal control strategy for CV systems using the Krotov algorithm to the covariance matrix's equation of motion to minimize the distance between the target CM and the controlled CM, facilitating precise numerical implementation without the need for Fock-basis cutoffs. Our proposed method allows for the study of complex quantum systems without suffering from the exponential growth of the Hilbert space. It is shown that, for a quadratic Hamiltonian, the corresponding coefficient matrix only grows as $2N$ where $N$ is the number of particles, compared to the exponential growth $N_c^N$ for a Fock-basis representation where $N_c$ is the cutoff per site. Our control strategy is also applicable to the case where a high-photon-number state is needed, or where the confinement of the dynamics in a small subspace becomes impossible. This precise, truncation-free macroscopic entanglement generation can have wide applications in novel quantum devices that use CV entangled states as a resource, such as quantum information processing devices~\cite{Weedbrook2012a,Braunstein2005a, Nielsen2000a} and quantum sensors~\cite{Cosco2021a,Weiss2021a}. Similar ideas may be applied to other CV systems or target states. For example, here we considered controlling the CM which encapsulates the first and second moments, but the algorithm may be easily generalized to control the dynamics of expectation values other than the first and second moments as long as they form a closed set of linear differential equations.

As an illustrative example, we applied the control strategy to generate quantum entanglement in an optomechanical system between a mechanical mirror and an optical cavity. Our control protocol reliably drive an initial vacuum state into an entangled two-mode squeezed state. Furthermore, the protocol exhibits good convergence behavior, with the distance between the controlled state and the target state converging exponentially with the number of Krotov iterations. The resulting control field is stable, exhibiting neither excessive amplitude nor high-frequency oscillations. We also imposed spectral constraints and found that satisfactory control fields can still be achieved within these limits. We also study the quantum speed limits and found the control is achieved when the runtime is larger than the limit imposed by the QSL. Finally, we have studied the effects of a general non-Markovian environment on the generated entanglement. The equation of motion for the CM under non-Markovian dynamics shows that quantum memory effects can preserve the entanglement, resulting in slower decay compared to a memoryless (Markovian) bath.
Our choice of the effective coupling rate $G=0.1 \omega_m$ could be feasible with current technologies~\cite{Aspelmeyer2014a, Dare2024a,Groblacher2009a,Rios-Sommer2021a}, and the protocol on the control of detuning could be a more experimentally viable choice~\cite{Milne2020a,Bishof2013a}. Note that theoretically, since detuning can be both positive and negative, it provides greater flexibility in calculating the control fields. In contrast, the laser intensity or coupling must be modulated to remain non-negative, which introduces an additional control constraint. Note in this case it is still possible to realize the desired control when the coupling is physically permissible to the control mechanism.
While our protocol allows the system to achieve the target entangled state within a predetermined time, it cannot preserve the target states indefinitely. One way to achieve long-term preservation of the target state is to have the ability to control the coupling $G$ (e.g., turn off or set the coupling $G$ to 0). In this case, the effective Hamiltonian would be $\omega_m b^\dagger  b - \Delta a^\dagger a$, which would merely add a phase to the squeezing parameter and keep the negativity (entanglement) constant after $t_f$.
The protocol and methodology presented here are general and adaptable to other Gaussian continuous variable systems~\cite{*[{After the completion of this work, we became aware of this recent work using similar techniques for charging of quantum batteries: }] [{}] Rodriguez2024a}, offering insights into optimal control strategies and the impact of open system dynamics.

\begin{acknowledgments}
	This work is supported in part by ACC-New Jersey under Contract No. W15QKN-18-D-0040.
\end{acknowledgments}

\onecolumngrid
\appendix
\section{The non-Markovian dynamics of the CM} \label{sec_appendix_dcm}

The details of deriving the equation of motion for the CM under a generic non-Markovian environment are presented here. For the noise-free $O$-operator considered here, the corresponding master equation for the reduced density operator of the system dynamics can be derived~\cite{Yu1999a} using the Novikov theorem for the ensemble average involving the noises, leading to
\begin{align}
	\frac{\partial}{\partial t} \rho_s(t)&=-i \left[H_s,\rho_t\right]+\left[L,\rho_t\bar{O}^{(0)\dagger}(t)\right]-\left[L ^\dagger,\bar{O}^{(0)}(t)\rho_t\right] \label{eq_append_meq}
\end{align}
With a system-bath coupling operator linear in $R$, $L=l_iR_i$, we may write down an ansatz for the $\bar{O}$-operator that's also only contain linear terms of $R_k$ as $\bar{O}(t)=o_i(t) R_i$. For brevity, we drop the explicit time-dependence of the $\bar{O}$ operator coefficients.

With the master equation Eq.~\eqref{eq_append_meq}, the Langevin equation for the first and second moments may be readily derived as
\begin{align}
	\partial_t \langle R_k \rangle = [\sigma\mathcal{M}]_{kj} \langle R_j \rangle + \left\{il_i o_j^*\sigma_{k,i} \langle R_j \rangle - i l_i^* o_j \sigma_{k,i} \langle R_j \rangle\right\}, \label{eq_cm1m}
\end{align}
and
\begin{align}
	\partial_t \langle R_iR_j \rangle
	&= \left[\sigma\mathcal{M}\right]_{ik} \langle R_k R_j\rangle + \langle R_i R_k \rangle \left[\sigma\mathcal{M}\right]^T_{kj} \nonumber \\
	& + \langle R_i R_n \rangle [io_n^* l_m \sigma^T_{mj} - io_n l_m^*\sigma^T_{mj} ] + [i\sigma_{im} l_m o_n^* - i\sigma_{im} l_m^* o_n] \langle R_n R_j \rangle \nonumber \\
	& + \sigma_{im} l_m^* o_n \sigma^T_{nj} + \sigma_{in} o_n^* l_m \sigma^T_{mj} \label{eq_cm2m}
\end{align}
Using the commutation relationship $\left[R_i, R_j \right] = i\sigma_{ij}$, we have
\begin{align}
	\langle R_iR_j \rangle + \langle R_jR_i \rangle
	= 2 \langle R_iR_j \rangle - i \sigma_{ij}.
\end{align}
Rearranging the terms in Eqs.~\eqref{eq_cm1m} and~\eqref{eq_cm2m} to plug into the CM elements $\gamma_{ij}=\langle R_i R_j \rangle+ \langle R_jR_i \rangle - 2\langle R_i \rangle \langle R_j \rangle$, it can then be shown that the CM follows
\begin{align}
	\partial_t \gamma &= \left[\sigma\mathcal{M} + \sigma \Delta \right] \gamma + \gamma \left[\sigma\mathcal{M} + \sigma \Delta\right]^T + 2 \left[\sigma \delta^R \sigma^T \right] \label{eq_append_dcm}
\end{align}
where $\delta_{mn} = l_m^* o_n + o_m^* l_n$, $\delta^R = \re [\delta]$, $\Delta_{mn}=i l_m o_n^* - i l_m^* o_n$.

Here, we will consider a leading order approximation of the $\bar{O}$ operator that keeps only the noise-independent terms,
\begin{align}
	\partial_t O(t,s) = \left[-iH_s - L^\dagger \bar{O}(t), O(t,s)\right]
\end{align}
which gives, with $\eta_{\rm eff} = \eta + i \Omega$,
\begin{align}
	\partial_t \bar{O} &= \alpha(0)L - \eta_{\rm eff}\bar{O} + \left[-iH_s - L ^\dagger \bar{O}, \bar{O}\right].
\end{align}
This allows us to get the differential equations for the $\bar{O}$ operator's coefficients as
\begin{align}
	\partial_t o_i
	&= \alpha(0) l_i - \eta_{\rm eff} o_i - o_l [\sigma\mathcal{M}]_{li} - i \sigma_{kl} \left[o_k o_l l^*_i  + o_i o_l l^*_k \right]
\end{align}
Together with Eq.~\eqref{eq_append_dcm}, this set of equations would allow one to directly calculate the dynamics of the CM under a non-Markovian bath for a generic quadratic Hamiltonian.

\twocolumngrid


\begin{thebibliography}{111}%
\makeatletter
\providecommand \@ifxundefined [1]{%
 \@ifx{#1\undefined}
}%
\providecommand \@ifnum [1]{%
 \ifnum #1\expandafter \@firstoftwo
 \else \expandafter \@secondoftwo
 \fi
}%
\providecommand \@ifx [1]{%
 \ifx #1\expandafter \@firstoftwo
 \else \expandafter \@secondoftwo
 \fi
}%
\providecommand \natexlab [1]{#1}%
\providecommand \enquote  [1]{``#1''}%
\providecommand \bibnamefont  [1]{#1}%
\providecommand \bibfnamefont [1]{#1}%
\providecommand \citenamefont [1]{#1}%
\providecommand \href@noop [0]{\@secondoftwo}%
\providecommand \href [0]{\begingroup \@sanitize@url \@href}%
\providecommand \@href[1]{\@@startlink{#1}\@@href}%
\providecommand \@@href[1]{\endgroup#1\@@endlink}%
\providecommand \@sanitize@url [0]{\catcode `\\12\catcode `\$12\catcode `\&12\catcode `\#12\catcode `\^12\catcode `\_12\catcode `\%12\relax}%
\providecommand \@@startlink[1]{}%
\providecommand \@@endlink[0]{}%
\providecommand \url  [0]{\begingroup\@sanitize@url \@url }%
\providecommand \@url [1]{\endgroup\@href {#1}{\urlprefix }}%
\providecommand \urlprefix  [0]{URL }%
\providecommand \Eprint [0]{\href }%
\providecommand \doibase [0]{https://doi.org/}%
\providecommand \selectlanguage [0]{\@gobble}%
\providecommand \bibinfo  [0]{\@secondoftwo}%
\providecommand \bibfield  [0]{\@secondoftwo}%
\providecommand \translation [1]{[#1]}%
\providecommand \BibitemOpen [0]{}%
\providecommand \bibitemStop [0]{}%
\providecommand \bibitemNoStop [0]{.\EOS\space}%
\providecommand \EOS [0]{\spacefactor3000\relax}%
\providecommand \BibitemShut  [1]{\csname bibitem#1\endcsname}%
\let\auto@bib@innerbib\@empty
\bibitem [{\citenamefont {Fr\"owis}\ \emph {et~al.}(2018)\citenamefont {Fr\"owis}, \citenamefont {Sekatski}, \citenamefont {D\"ur}, \citenamefont {Gisin},\ and\ \citenamefont {Sangouard}}]{Frowis2018a}%
  \BibitemOpen
  \bibfield  {author} {\bibinfo {author} {\bibfnamefont {F.}~\bibnamefont {Fr\"owis}}, \bibinfo {author} {\bibfnamefont {P.}~\bibnamefont {Sekatski}}, \bibinfo {author} {\bibfnamefont {W.}~\bibnamefont {D\"ur}}, \bibinfo {author} {\bibfnamefont {N.}~\bibnamefont {Gisin}},\ and\ \bibinfo {author} {\bibfnamefont {N.}~\bibnamefont {Sangouard}},\ }\bibfield  {title} {\bibinfo {title} {{Macroscopic quantum states: Measures, fragility, and implementations}},\ }\href {https://doi.org/10.1103/RevModPhys.90.025004} {\bibfield  {journal} {\bibinfo  {journal} {Rev. Mod. Phys.}\ }\textbf {\bibinfo {volume} {90}},\ \bibinfo {pages} {025004} (\bibinfo {year} {2018})}\BibitemShut {NoStop}%
\bibitem [{\citenamefont {Leggett}(1980)}]{Leggett1980a}%
  \BibitemOpen
  \bibfield  {author} {\bibinfo {author} {\bibfnamefont {A.~J.}\ \bibnamefont {Leggett}},\ }\bibfield  {title} {\bibinfo {title} {{{Macroscopic Quantum Systems and the Quantum Theory of Measurement}}},\ }\href {https://doi.org/10.1143/PTP.69.80} {\bibfield  {journal} {\bibinfo  {journal} {Progress of Theoretical Physics Supplement}\ }\textbf {\bibinfo {volume} {69}},\ \bibinfo {pages} {80} (\bibinfo {year} {1980})}\BibitemShut {NoStop}%
\bibitem [{\citenamefont {Farrow}\ and\ \citenamefont {Vedral}(2015)}]{Farrow2015a}%
  \BibitemOpen
  \bibfield  {author} {\bibinfo {author} {\bibfnamefont {T.}~\bibnamefont {Farrow}}\ and\ \bibinfo {author} {\bibfnamefont {V.}~\bibnamefont {Vedral}},\ }\bibfield  {title} {\bibinfo {title} {{Classification of macroscopic quantum effects}},\ }\href {https://doi.org/https://doi.org/10.1016/j.optcom.2014.06.042} {\bibfield  {journal} {\bibinfo  {journal} {Optics Communications}\ }\textbf {\bibinfo {volume} {337}},\ \bibinfo {pages} {22} (\bibinfo {year} {2015})},\ \bibinfo {note} {macroscopic quantumness: theory and applications in optical sciences}\BibitemShut {NoStop}%
\bibitem [{\citenamefont {Caldeira}(2014)}]{Caldeira2014a}%
  \BibitemOpen
  \bibfield  {author} {\bibinfo {author} {\bibfnamefont {A.~O.}\ \bibnamefont {Caldeira}},\ }\href@noop {} {\emph {\bibinfo {title} {{An introduction to macroscopic quantum phenomena and quantum dissipation}}}}\ (\bibinfo  {publisher} {Cambridge University Press},\ \bibinfo {year} {2014})\BibitemShut {NoStop}%
\bibitem [{\citenamefont {Lachance-Quirion}\ \emph {et~al.}(2017)\citenamefont {Lachance-Quirion}, \citenamefont {Tabuchi}, \citenamefont {Ishino}, \citenamefont {Noguchi}, \citenamefont {Ishikawa}, \citenamefont {Yamazaki},\ and\ \citenamefont {Nakamura}}]{Lachance-Quirion2017a}%
  \BibitemOpen
  \bibfield  {author} {\bibinfo {author} {\bibfnamefont {D.}~\bibnamefont {Lachance-Quirion}}, \bibinfo {author} {\bibfnamefont {Y.}~\bibnamefont {Tabuchi}}, \bibinfo {author} {\bibfnamefont {S.}~\bibnamefont {Ishino}}, \bibinfo {author} {\bibfnamefont {A.}~\bibnamefont {Noguchi}}, \bibinfo {author} {\bibfnamefont {T.}~\bibnamefont {Ishikawa}}, \bibinfo {author} {\bibfnamefont {R.}~\bibnamefont {Yamazaki}},\ and\ \bibinfo {author} {\bibfnamefont {Y.}~\bibnamefont {Nakamura}},\ }\bibfield  {title} {\bibinfo {title} {{Resolving quanta of collective spin excitations in a millimeter-sized ferromagnet}},\ }\bibfield  {journal} {\bibinfo  {journal} {Science Advances}\ }\textbf {\bibinfo {volume} {3}},\ \href {https://doi.org/10.1126/sciadv.1603150} {10.1126/sciadv.1603150} (\bibinfo {year} {2017})\BibitemShut {NoStop}%
\bibitem [{\citenamefont {Li}\ \emph {et~al.}(2018)\citenamefont {Li}, \citenamefont {Zhu},\ and\ \citenamefont {Agarwal}}]{Li2018a}%
  \BibitemOpen
  \bibfield  {author} {\bibinfo {author} {\bibfnamefont {J.}~\bibnamefont {Li}}, \bibinfo {author} {\bibfnamefont {S.-Y.}\ \bibnamefont {Zhu}},\ and\ \bibinfo {author} {\bibfnamefont {G.~S.}\ \bibnamefont {Agarwal}},\ }\bibfield  {title} {\bibinfo {title} {{Magnon-Photon-Phonon Entanglement in Cavity Magnomechanics}},\ }\href {https://doi.org/10.1103/physrevlett.121.203601} {\bibfield  {journal} {\bibinfo  {journal} {Phys. Rev. Lett.}\ }\textbf {\bibinfo {volume} {121}},\ \bibinfo {pages} {203601} (\bibinfo {year} {2018})}\BibitemShut {NoStop}%
\bibitem [{\citenamefont {Zhang}\ \emph {et~al.}(2015)\citenamefont {Zhang}, \citenamefont {Wang}, \citenamefont {Li}, \citenamefont {Luo}, \citenamefont {Wu}, \citenamefont {Nori},\ and\ \citenamefont {You}}]{you_nat1}%
  \BibitemOpen
  \bibfield  {author} {\bibinfo {author} {\bibfnamefont {D.}~\bibnamefont {Zhang}}, \bibinfo {author} {\bibfnamefont {X.-M.}\ \bibnamefont {Wang}}, \bibinfo {author} {\bibfnamefont {T.-F.}\ \bibnamefont {Li}}, \bibinfo {author} {\bibfnamefont {X.-Q.}\ \bibnamefont {Luo}}, \bibinfo {author} {\bibfnamefont {W.}~\bibnamefont {Wu}}, \bibinfo {author} {\bibfnamefont {F.}~\bibnamefont {Nori}},\ and\ \bibinfo {author} {\bibfnamefont {J.}~\bibnamefont {You}},\ }\bibfield  {title} {\bibinfo {title} {{Cavity quantum electrodynamics with ferromagnetic magnons in a small yttrium-iron-garnet sphere}},\ }\href {https://doi.org/10.1038/npjqi.2015.14} {\bibfield  {journal} {\bibinfo  {journal} {npj Quantum Information}\ }\textbf {\bibinfo {volume} {1}},\ \bibinfo {pages} {15014} (\bibinfo {year} {2015})}\BibitemShut {NoStop}%
\bibitem [{\citenamefont {Zhang}\ \emph {et~al.}(2016)\citenamefont {Zhang}, \citenamefont {Zou}, \citenamefont {Jiang},\ and\ \citenamefont {Tang}}]{Zhang2016a}%
  \BibitemOpen
  \bibfield  {author} {\bibinfo {author} {\bibfnamefont {X.}~\bibnamefont {Zhang}}, \bibinfo {author} {\bibfnamefont {C.-L.}\ \bibnamefont {Zou}}, \bibinfo {author} {\bibfnamefont {L.}~\bibnamefont {Jiang}},\ and\ \bibinfo {author} {\bibfnamefont {H.~X.}\ \bibnamefont {Tang}},\ }\bibfield  {title} {\bibinfo {title} {{Cavity magnomechanics}},\ }\href {https://doi.org/10.1126/sciadv.1501286} {\bibfield  {journal} {\bibinfo  {journal} {Science Advances}\ }\textbf {\bibinfo {volume} {2}},\ \bibinfo {pages} {e1501286} (\bibinfo {year} {2016})}\BibitemShut {NoStop}%
\bibitem [{\citenamefont {Xu}\ \emph {et~al.}(2023)\citenamefont {Xu}, \citenamefont {Gu}, \citenamefont {Li}, \citenamefont {Weng}, \citenamefont {Wang}, \citenamefont {Li}, \citenamefont {Wang}, \citenamefont {Zhu},\ and\ \citenamefont {You}}]{Xu2023a}%
  \BibitemOpen
  \bibfield  {author} {\bibinfo {author} {\bibfnamefont {D.}~\bibnamefont {Xu}}, \bibinfo {author} {\bibfnamefont {X.-K.}\ \bibnamefont {Gu}}, \bibinfo {author} {\bibfnamefont {H.-K.}\ \bibnamefont {Li}}, \bibinfo {author} {\bibfnamefont {Y.-C.}\ \bibnamefont {Weng}}, \bibinfo {author} {\bibfnamefont {Y.-P.}\ \bibnamefont {Wang}}, \bibinfo {author} {\bibfnamefont {J.}~\bibnamefont {Li}}, \bibinfo {author} {\bibfnamefont {H.}~\bibnamefont {Wang}}, \bibinfo {author} {\bibfnamefont {S.-Y.}\ \bibnamefont {Zhu}},\ and\ \bibinfo {author} {\bibfnamefont {J.~Q.}\ \bibnamefont {You}},\ }\bibfield  {title} {\bibinfo {title} {{Quantum Control of a Single Magnon in a Macroscopic Spin System}},\ }\href {https://doi.org/10.1103/PhysRevLett.130.193603} {\bibfield  {journal} {\bibinfo  {journal} {Phys. Rev. Lett.}\ }\textbf {\bibinfo {volume} {130}},\ \bibinfo {pages} {193603} (\bibinfo {year} {2023})}\BibitemShut {NoStop}%
\bibitem [{\citenamefont {Luo}\ \emph {et~al.}(2021)\citenamefont {Luo}, \citenamefont {Qian},\ and\ \citenamefont {Yu}}]{Luo2021a}%
  \BibitemOpen
  \bibfield  {author} {\bibinfo {author} {\bibfnamefont {D.-W.}\ \bibnamefont {Luo}}, \bibinfo {author} {\bibfnamefont {X.-F.}\ \bibnamefont {Qian}},\ and\ \bibinfo {author} {\bibfnamefont {T.}~\bibnamefont {Yu}},\ }\bibfield  {title} {\bibinfo {title} {{Nonlocal magnon entanglement generation in coupled hybrid cavity systems}},\ }\href {https://doi.org/10.1364/OL.414975} {\bibfield  {journal} {\bibinfo  {journal} {Opt. Lett.}\ }\textbf {\bibinfo {volume} {46}},\ \bibinfo {pages} {1073} (\bibinfo {year} {2021})}\BibitemShut {NoStop}%
\bibitem [{\citenamefont {Luo}\ \emph {et~al.}(2024)\citenamefont {Luo}, \citenamefont {Qian},\ and\ \citenamefont {Yu}}]{Luo2024a}%
  \BibitemOpen
  \bibfield  {author} {\bibinfo {author} {\bibfnamefont {D.-W.}\ \bibnamefont {Luo}}, \bibinfo {author} {\bibfnamefont {X.-F.}\ \bibnamefont {Qian}},\ and\ \bibinfo {author} {\bibfnamefont {T.}~\bibnamefont {Yu}},\ }\bibfield  {title} {\bibinfo {title} {{Optically mediated remote entanglement generation in magnon-cavity systems}},\ }\href {https://doi.org/10.1103/PhysRevA.109.012611} {\bibfield  {journal} {\bibinfo  {journal} {Phys. Rev. A}\ }\textbf {\bibinfo {volume} {109}},\ \bibinfo {pages} {012611} (\bibinfo {year} {2024})}\BibitemShut {NoStop}%
\bibitem [{\citenamefont {Markus~Aspelmeyer}(2014)}]{cavoptmbook2014s}%
  \BibitemOpen
  \bibfield  {author} {\bibinfo {author} {\bibfnamefont {F.~M.}\ \bibnamefont {Markus~Aspelmeyer}, \bibfnamefont {Tobias J.~Kippenberg}},\ }\href {https://doi.org/10.1007/978-3-642-55312-7} {\emph {\bibinfo {title} {{Cavity Optomechanics: Nano- and Micromechanical Resonators Interacting with Light}}}}\ (\bibinfo  {publisher} {Springer Berlin Heidelberg},\ \bibinfo {year} {2014})\BibitemShut {NoStop}%
\bibitem [{\citenamefont {Aspelmeyer}\ \emph {et~al.}(2014)\citenamefont {Aspelmeyer}, \citenamefont {Kippenberg},\ and\ \citenamefont {Marquardt}}]{Aspelmeyer2014a}%
  \BibitemOpen
  \bibfield  {author} {\bibinfo {author} {\bibfnamefont {M.}~\bibnamefont {Aspelmeyer}}, \bibinfo {author} {\bibfnamefont {T.~J.}\ \bibnamefont {Kippenberg}},\ and\ \bibinfo {author} {\bibfnamefont {F.}~\bibnamefont {Marquardt}},\ }\bibfield  {title} {\bibinfo {title} {{Cavity optomechanics}},\ }\href {https://doi.org/10.1103/RevModPhys.86.1391} {\bibfield  {journal} {\bibinfo  {journal} {Rev. Mod. Phys.}\ }\textbf {\bibinfo {volume} {86}},\ \bibinfo {pages} {1391} (\bibinfo {year} {2014})}\BibitemShut {NoStop}%
\bibitem [{\citenamefont {Barzanjeh}\ \emph {et~al.}(2022)\citenamefont {Barzanjeh}, \citenamefont {Xuereb}, \citenamefont {Gr{\"o}blacher}, \citenamefont {Paternostro}, \citenamefont {Regal},\ and\ \citenamefont {Weig}}]{Barzanjeh2022a}%
  \BibitemOpen
  \bibfield  {author} {\bibinfo {author} {\bibfnamefont {S.}~\bibnamefont {Barzanjeh}}, \bibinfo {author} {\bibfnamefont {A.}~\bibnamefont {Xuereb}}, \bibinfo {author} {\bibfnamefont {S.}~\bibnamefont {Gr{\"o}blacher}}, \bibinfo {author} {\bibfnamefont {M.}~\bibnamefont {Paternostro}}, \bibinfo {author} {\bibfnamefont {C.~A.}\ \bibnamefont {Regal}},\ and\ \bibinfo {author} {\bibfnamefont {E.~M.}\ \bibnamefont {Weig}},\ }\bibfield  {title} {\bibinfo {title} {{Optomechanics for quantum technologies}},\ }\href@noop {} {\bibfield  {journal} {\bibinfo  {journal} {Nature Physics}\ }\textbf {\bibinfo {volume} {18}},\ \bibinfo {pages} {15} (\bibinfo {year} {2022})}\BibitemShut {NoStop}%
\bibitem [{\citenamefont {Meystre}(2013)}]{Meystre2013a}%
  \BibitemOpen
  \bibfield  {author} {\bibinfo {author} {\bibfnamefont {P.}~\bibnamefont {Meystre}},\ }\bibfield  {title} {\bibinfo {title} {{A short walk through quantum optomechanics}},\ }\href@noop {} {\bibfield  {journal} {\bibinfo  {journal} {Annalen der Physik}\ }\textbf {\bibinfo {volume} {525}},\ \bibinfo {pages} {215} (\bibinfo {year} {2013})}\BibitemShut {NoStop}%
\bibitem [{\citenamefont {Milburn}\ and\ \citenamefont {Woolley}(2011)}]{Milburn2011a}%
  \BibitemOpen
  \bibfield  {author} {\bibinfo {author} {\bibfnamefont {G.}~\bibnamefont {Milburn}}\ and\ \bibinfo {author} {\bibfnamefont {M.}~\bibnamefont {Woolley}},\ }\bibfield  {title} {\bibinfo {title} {{An introduction to quantum optomechanics}},\ }\href@noop {} {\bibfield  {journal} {\bibinfo  {journal} {acta physica slovaca}\ }\textbf {\bibinfo {volume} {61}},\ \bibinfo {pages} {483} (\bibinfo {year} {2011})}\BibitemShut {NoStop}%
\bibitem [{\citenamefont {Rashid}\ \emph {et~al.}(2016)\citenamefont {Rashid}, \citenamefont {Tufarelli}, \citenamefont {Bateman}, \citenamefont {Vovrosh}, \citenamefont {Hempston}, \citenamefont {Kim},\ and\ \citenamefont {Ulbricht}}]{Rashid2016a}%
  \BibitemOpen
  \bibfield  {author} {\bibinfo {author} {\bibfnamefont {M.}~\bibnamefont {Rashid}}, \bibinfo {author} {\bibfnamefont {T.}~\bibnamefont {Tufarelli}}, \bibinfo {author} {\bibfnamefont {J.}~\bibnamefont {Bateman}}, \bibinfo {author} {\bibfnamefont {J.}~\bibnamefont {Vovrosh}}, \bibinfo {author} {\bibfnamefont {D.}~\bibnamefont {Hempston}}, \bibinfo {author} {\bibfnamefont {M.~S.}\ \bibnamefont {Kim}},\ and\ \bibinfo {author} {\bibfnamefont {H.}~\bibnamefont {Ulbricht}},\ }\bibfield  {title} {\bibinfo {title} {{Experimental Realization of a Thermal Squeezed State of Levitated Optomechanics}},\ }\href {https://doi.org/10.1103/PhysRevLett.117.273601} {\bibfield  {journal} {\bibinfo  {journal} {Phys. Rev. Lett.}\ }\textbf {\bibinfo {volume} {117}},\ \bibinfo {pages} {273601} (\bibinfo {year} {2016})}\BibitemShut {NoStop}%
\bibitem [{\citenamefont {Hoang}\ \emph {et~al.}(2016)\citenamefont {Hoang}, \citenamefont {Ma}, \citenamefont {Ahn}, \citenamefont {Bang}, \citenamefont {Robicheaux}, \citenamefont {Yin},\ and\ \citenamefont {Li}}]{Hoang2016a}%
  \BibitemOpen
  \bibfield  {author} {\bibinfo {author} {\bibfnamefont {T.~M.}\ \bibnamefont {Hoang}}, \bibinfo {author} {\bibfnamefont {Y.}~\bibnamefont {Ma}}, \bibinfo {author} {\bibfnamefont {J.}~\bibnamefont {Ahn}}, \bibinfo {author} {\bibfnamefont {J.}~\bibnamefont {Bang}}, \bibinfo {author} {\bibfnamefont {F.}~\bibnamefont {Robicheaux}}, \bibinfo {author} {\bibfnamefont {Z.-Q.}\ \bibnamefont {Yin}},\ and\ \bibinfo {author} {\bibfnamefont {T.}~\bibnamefont {Li}},\ }\bibfield  {title} {\bibinfo {title} {{Torsional Optomechanics of a Levitated Nonspherical Nanoparticle}},\ }\href {https://doi.org/10.1103/PhysRevLett.117.123604} {\bibfield  {journal} {\bibinfo  {journal} {Phys. Rev. Lett.}\ }\textbf {\bibinfo {volume} {117}},\ \bibinfo {pages} {123604} (\bibinfo {year} {2016})}\BibitemShut {NoStop}%
\bibitem [{\citenamefont {Dobrindt}\ \emph {et~al.}(2008)\citenamefont {Dobrindt}, \citenamefont {Wilson-Rae},\ and\ \citenamefont {Kippenberg}}]{Dobrindt2008a}%
  \BibitemOpen
  \bibfield  {author} {\bibinfo {author} {\bibfnamefont {J.~M.}\ \bibnamefont {Dobrindt}}, \bibinfo {author} {\bibfnamefont {I.}~\bibnamefont {Wilson-Rae}},\ and\ \bibinfo {author} {\bibfnamefont {T.~J.}\ \bibnamefont {Kippenberg}},\ }\bibfield  {title} {\bibinfo {title} {{Parametric Normal-Mode Splitting in Cavity Optomechanics}},\ }\href {https://doi.org/10.1103/PhysRevLett.101.263602} {\bibfield  {journal} {\bibinfo  {journal} {Phys. Rev. Lett.}\ }\textbf {\bibinfo {volume} {101}},\ \bibinfo {pages} {263602} (\bibinfo {year} {2008})}\BibitemShut {NoStop}%
\bibitem [{\citenamefont {Stannigel}\ \emph {et~al.}(2010)\citenamefont {Stannigel}, \citenamefont {Rabl}, \citenamefont {S\o{}rensen}, \citenamefont {Zoller},\ and\ \citenamefont {Lukin}}]{Stannigel2010a}%
  \BibitemOpen
  \bibfield  {author} {\bibinfo {author} {\bibfnamefont {K.}~\bibnamefont {Stannigel}}, \bibinfo {author} {\bibfnamefont {P.}~\bibnamefont {Rabl}}, \bibinfo {author} {\bibfnamefont {A.~S.}\ \bibnamefont {S\o{}rensen}}, \bibinfo {author} {\bibfnamefont {P.}~\bibnamefont {Zoller}},\ and\ \bibinfo {author} {\bibfnamefont {M.~D.}\ \bibnamefont {Lukin}},\ }\bibfield  {title} {\bibinfo {title} {{Optomechanical Transducers for Long-Distance Quantum Communication}},\ }\href {https://doi.org/10.1103/PhysRevLett.105.220501} {\bibfield  {journal} {\bibinfo  {journal} {Phys. Rev. Lett.}\ }\textbf {\bibinfo {volume} {105}},\ \bibinfo {pages} {220501} (\bibinfo {year} {2010})}\BibitemShut {NoStop}%
\bibitem [{\citenamefont {Fiore}\ \emph {et~al.}(2011)\citenamefont {Fiore}, \citenamefont {Yang}, \citenamefont {Kuzyk}, \citenamefont {Barbour}, \citenamefont {Tian},\ and\ \citenamefont {Wang}}]{Fiore2011a}%
  \BibitemOpen
  \bibfield  {author} {\bibinfo {author} {\bibfnamefont {V.}~\bibnamefont {Fiore}}, \bibinfo {author} {\bibfnamefont {Y.}~\bibnamefont {Yang}}, \bibinfo {author} {\bibfnamefont {M.~C.}\ \bibnamefont {Kuzyk}}, \bibinfo {author} {\bibfnamefont {R.}~\bibnamefont {Barbour}}, \bibinfo {author} {\bibfnamefont {L.}~\bibnamefont {Tian}},\ and\ \bibinfo {author} {\bibfnamefont {H.}~\bibnamefont {Wang}},\ }\bibfield  {title} {\bibinfo {title} {{Storing Optical Information as a Mechanical Excitation in a Silica Optomechanical Resonator}},\ }\href {https://doi.org/10.1103/PhysRevLett.107.133601} {\bibfield  {journal} {\bibinfo  {journal} {Phys. Rev. Lett.}\ }\textbf {\bibinfo {volume} {107}},\ \bibinfo {pages} {133601} (\bibinfo {year} {2011})}\BibitemShut {NoStop}%
\bibitem [{\citenamefont {Stannigel}\ \emph {et~al.}(2012)\citenamefont {Stannigel}, \citenamefont {Komar}, \citenamefont {Habraken}, \citenamefont {Bennett}, \citenamefont {Lukin}, \citenamefont {Zoller},\ and\ \citenamefont {Rabl}}]{Stannigel2012a}%
  \BibitemOpen
  \bibfield  {author} {\bibinfo {author} {\bibfnamefont {K.}~\bibnamefont {Stannigel}}, \bibinfo {author} {\bibfnamefont {P.}~\bibnamefont {Komar}}, \bibinfo {author} {\bibfnamefont {S.~J.~M.}\ \bibnamefont {Habraken}}, \bibinfo {author} {\bibfnamefont {S.~D.}\ \bibnamefont {Bennett}}, \bibinfo {author} {\bibfnamefont {M.~D.}\ \bibnamefont {Lukin}}, \bibinfo {author} {\bibfnamefont {P.}~\bibnamefont {Zoller}},\ and\ \bibinfo {author} {\bibfnamefont {P.}~\bibnamefont {Rabl}},\ }\bibfield  {title} {\bibinfo {title} {{Optomechanical Quantum Information Processing with Photons and Phonons}},\ }\href {https://doi.org/10.1103/PhysRevLett.109.013603} {\bibfield  {journal} {\bibinfo  {journal} {Phys. Rev. Lett.}\ }\textbf {\bibinfo {volume} {109}},\ \bibinfo {pages} {013603} (\bibinfo {year} {2012})}\BibitemShut {NoStop}%
\bibitem [{\citenamefont {Stannigel}\ \emph {et~al.}(2011)\citenamefont {Stannigel}, \citenamefont {Rabl}, \citenamefont {S\o{}rensen}, \citenamefont {Lukin},\ and\ \citenamefont {Zoller}}]{Stannigel2011a}%
  \BibitemOpen
  \bibfield  {author} {\bibinfo {author} {\bibfnamefont {K.}~\bibnamefont {Stannigel}}, \bibinfo {author} {\bibfnamefont {P.}~\bibnamefont {Rabl}}, \bibinfo {author} {\bibfnamefont {A.~S.}\ \bibnamefont {S\o{}rensen}}, \bibinfo {author} {\bibfnamefont {M.~D.}\ \bibnamefont {Lukin}},\ and\ \bibinfo {author} {\bibfnamefont {P.}~\bibnamefont {Zoller}},\ }\bibfield  {title} {\bibinfo {title} {{Optomechanical transducers for quantum-information processing}},\ }\href {https://doi.org/10.1103/PhysRevA.84.042341} {\bibfield  {journal} {\bibinfo  {journal} {Phys. Rev. A}\ }\textbf {\bibinfo {volume} {84}},\ \bibinfo {pages} {042341} (\bibinfo {year} {2011})}\BibitemShut {NoStop}%
\bibitem [{\citenamefont {Yang}\ \emph {et~al.}(2015)\citenamefont {Yang}, \citenamefont {Zhang}, \citenamefont {Wang}, \citenamefont {Liu}, \citenamefont {Wu}, \citenamefont {Liu}, \citenamefont {Li},\ and\ \citenamefont {Nori}}]{yang2015noise}%
  \BibitemOpen
  \bibfield  {author} {\bibinfo {author} {\bibfnamefont {N.}~\bibnamefont {Yang}}, \bibinfo {author} {\bibfnamefont {J.}~\bibnamefont {Zhang}}, \bibinfo {author} {\bibfnamefont {H.}~\bibnamefont {Wang}}, \bibinfo {author} {\bibfnamefont {Y.~X.}\ \bibnamefont {Liu}}, \bibinfo {author} {\bibfnamefont {R.~B.}\ \bibnamefont {Wu}}, \bibinfo {author} {\bibfnamefont {L.~Q.}\ \bibnamefont {Liu}}, \bibinfo {author} {\bibfnamefont {C.~W.}\ \bibnamefont {Li}},\ and\ \bibinfo {author} {\bibfnamefont {F.}~\bibnamefont {Nori}},\ }\bibfield  {title} {\bibinfo {title} {{Noise suppression of on-chip mechanical resonators by chaotic coherent feedback}},\ }\href@noop {} {\bibfield  {journal} {\bibinfo  {journal} {Phys. Rev. A}\ }\textbf {\bibinfo {volume} {92}},\ \bibinfo {pages} {033812} (\bibinfo {year} {2015})}\BibitemShut {NoStop}%
\bibitem [{\citenamefont {Jiang}\ \emph {et~al.}(2017)\citenamefont {Jiang}, \citenamefont {Shao}, \citenamefont {Zhang}, \citenamefont {Yi}, \citenamefont {Wiersig}, \citenamefont {Wang}, \citenamefont {Gong}, \citenamefont {Loncar}, \citenamefont {Yang},\ and\ \citenamefont {Xiao}}]{jiang2017chaos}%
  \BibitemOpen
  \bibfield  {author} {\bibinfo {author} {\bibfnamefont {X.~F.}\ \bibnamefont {Jiang}}, \bibinfo {author} {\bibfnamefont {L.~B.}\ \bibnamefont {Shao}}, \bibinfo {author} {\bibfnamefont {S.~X.}\ \bibnamefont {Zhang}}, \bibinfo {author} {\bibfnamefont {X.}~\bibnamefont {Yi}}, \bibinfo {author} {\bibfnamefont {J.}~\bibnamefont {Wiersig}}, \bibinfo {author} {\bibfnamefont {L.}~\bibnamefont {Wang}}, \bibinfo {author} {\bibfnamefont {Q.~H.}\ \bibnamefont {Gong}}, \bibinfo {author} {\bibfnamefont {M.}~\bibnamefont {Loncar}}, \bibinfo {author} {\bibfnamefont {L.}~\bibnamefont {Yang}},\ and\ \bibinfo {author} {\bibfnamefont {Y.~F.}\ \bibnamefont {Xiao}},\ }\bibfield  {title} {\bibinfo {title} {{Chaos-assisted broadband momentum transformation in optical microresonators}},\ }\href@noop {} {\bibfield  {journal} {\bibinfo  {journal} {Science}\ }\textbf {\bibinfo {volume} {358}},\ \bibinfo {pages} {344} (\bibinfo {year} {2017})}\BibitemShut {NoStop}%
\bibitem [{\citenamefont {Mu}\ \emph {et~al.}(2016)\citenamefont {Mu}, \citenamefont {Zhao},\ and\ \citenamefont {Yu}}]{mu2016memory}%
  \BibitemOpen
  \bibfield  {author} {\bibinfo {author} {\bibfnamefont {Q.}~\bibnamefont {Mu}}, \bibinfo {author} {\bibfnamefont {X.}~\bibnamefont {Zhao}},\ and\ \bibinfo {author} {\bibfnamefont {T.}~\bibnamefont {Yu}},\ }\bibfield  {title} {\bibinfo {title} {{Memory-effect-induced macroscopic-microscopic entanglement}},\ }\href@noop {} {\bibfield  {journal} {\bibinfo  {journal} {Phys. Rev. A}\ }\textbf {\bibinfo {volume} {94}},\ \bibinfo {pages} {012334} (\bibinfo {year} {2016})}\BibitemShut {NoStop}%
\bibitem [{\citenamefont {Liao}\ \emph {et~al.}(2014)\citenamefont {Liao}, \citenamefont {Wu},\ and\ \citenamefont {Nori}}]{Liao2014a}%
  \BibitemOpen
  \bibfield  {author} {\bibinfo {author} {\bibfnamefont {J.-Q.}\ \bibnamefont {Liao}}, \bibinfo {author} {\bibfnamefont {Q.-Q.}\ \bibnamefont {Wu}},\ and\ \bibinfo {author} {\bibfnamefont {F.}~\bibnamefont {Nori}},\ }\bibfield  {title} {\bibinfo {title} {{Entangling two macroscopic mechanical mirrors in a two-cavity optomechanical system}},\ }\href {https://doi.org/10.1103/PhysRevA.89.014302} {\bibfield  {journal} {\bibinfo  {journal} {Phys. Rev. A}\ }\textbf {\bibinfo {volume} {89}},\ \bibinfo {pages} {014302} (\bibinfo {year} {2014})}\BibitemShut {NoStop}%
\bibitem [{\citenamefont {Wang}\ \emph {et~al.}(2016)\citenamefont {Wang}, \citenamefont {L\"u}, \citenamefont {Wang}, \citenamefont {You},\ and\ \citenamefont {Wu}}]{Wang2016a}%
  \BibitemOpen
  \bibfield  {author} {\bibinfo {author} {\bibfnamefont {M.}~\bibnamefont {Wang}}, \bibinfo {author} {\bibfnamefont {X.-Y.}\ \bibnamefont {L\"u}}, \bibinfo {author} {\bibfnamefont {Y.-D.}\ \bibnamefont {Wang}}, \bibinfo {author} {\bibfnamefont {J.~Q.}\ \bibnamefont {You}},\ and\ \bibinfo {author} {\bibfnamefont {Y.}~\bibnamefont {Wu}},\ }\bibfield  {title} {\bibinfo {title} {{Macroscopic quantum entanglement in modulated optomechanics}},\ }\href {https://doi.org/10.1103/PhysRevA.94.053807} {\bibfield  {journal} {\bibinfo  {journal} {Phys. Rev. A}\ }\textbf {\bibinfo {volume} {94}},\ \bibinfo {pages} {053807} (\bibinfo {year} {2016})}\BibitemShut {NoStop}%
\bibitem [{\citenamefont {Vitali}\ \emph {et~al.}(2007)\citenamefont {Vitali}, \citenamefont {Gigan}, \citenamefont {Ferreira}, \citenamefont {B\"ohm}, \citenamefont {Tombesi}, \citenamefont {Guerreiro}, \citenamefont {Vedral}, \citenamefont {Zeilinger},\ and\ \citenamefont {Aspelmeyer}}]{Vitali2007a}%
  \BibitemOpen
  \bibfield  {author} {\bibinfo {author} {\bibfnamefont {D.}~\bibnamefont {Vitali}}, \bibinfo {author} {\bibfnamefont {S.}~\bibnamefont {Gigan}}, \bibinfo {author} {\bibfnamefont {A.}~\bibnamefont {Ferreira}}, \bibinfo {author} {\bibfnamefont {H.~R.}\ \bibnamefont {B\"ohm}}, \bibinfo {author} {\bibfnamefont {P.}~\bibnamefont {Tombesi}}, \bibinfo {author} {\bibfnamefont {A.}~\bibnamefont {Guerreiro}}, \bibinfo {author} {\bibfnamefont {V.}~\bibnamefont {Vedral}}, \bibinfo {author} {\bibfnamefont {A.}~\bibnamefont {Zeilinger}},\ and\ \bibinfo {author} {\bibfnamefont {M.}~\bibnamefont {Aspelmeyer}},\ }\bibfield  {title} {\bibinfo {title} {{Optomechanical Entanglement between a Movable Mirror and a Cavity Field}},\ }\href {https://doi.org/10.1103/PhysRevLett.98.030405} {\bibfield  {journal} {\bibinfo  {journal} {Phys. Rev. Lett.}\ }\textbf {\bibinfo {volume} {98}},\ \bibinfo {pages} {030405} (\bibinfo {year} {2007})}\BibitemShut {NoStop}%
\bibitem [{\citenamefont {Mancini}\ \emph {et~al.}(2002)\citenamefont {Mancini}, \citenamefont {Giovannetti}, \citenamefont {Vitali},\ and\ \citenamefont {Tombesi}}]{Mancini2002a}%
  \BibitemOpen
  \bibfield  {author} {\bibinfo {author} {\bibfnamefont {S.}~\bibnamefont {Mancini}}, \bibinfo {author} {\bibfnamefont {V.}~\bibnamefont {Giovannetti}}, \bibinfo {author} {\bibfnamefont {D.}~\bibnamefont {Vitali}},\ and\ \bibinfo {author} {\bibfnamefont {P.}~\bibnamefont {Tombesi}},\ }\bibfield  {title} {\bibinfo {title} {{Entangling Macroscopic Oscillators Exploiting Radiation Pressure}},\ }\href {https://doi.org/10.1103/PhysRevLett.88.120401} {\bibfield  {journal} {\bibinfo  {journal} {Phys. Rev. Lett.}\ }\textbf {\bibinfo {volume} {88}},\ \bibinfo {pages} {120401} (\bibinfo {year} {2002})}\BibitemShut {NoStop}%
\bibitem [{\citenamefont {Tittonen}\ \emph {et~al.}(1999)\citenamefont {Tittonen}, \citenamefont {Breitenbach}, \citenamefont {Kalkbrenner}, \citenamefont {M\"uller}, \citenamefont {Conradt}, \citenamefont {Schiller}, \citenamefont {Steinsland}, \citenamefont {Blanc},\ and\ \citenamefont {de~Rooij}}]{Tittonen1999a}%
  \BibitemOpen
  \bibfield  {author} {\bibinfo {author} {\bibfnamefont {I.}~\bibnamefont {Tittonen}}, \bibinfo {author} {\bibfnamefont {G.}~\bibnamefont {Breitenbach}}, \bibinfo {author} {\bibfnamefont {T.}~\bibnamefont {Kalkbrenner}}, \bibinfo {author} {\bibfnamefont {T.}~\bibnamefont {M\"uller}}, \bibinfo {author} {\bibfnamefont {R.}~\bibnamefont {Conradt}}, \bibinfo {author} {\bibfnamefont {S.}~\bibnamefont {Schiller}}, \bibinfo {author} {\bibfnamefont {E.}~\bibnamefont {Steinsland}}, \bibinfo {author} {\bibfnamefont {N.}~\bibnamefont {Blanc}},\ and\ \bibinfo {author} {\bibfnamefont {N.~F.}\ \bibnamefont {de~Rooij}},\ }\bibfield  {title} {\bibinfo {title} {{Interferometric measurements of the position of a macroscopic body: Towards observation of quantum limits}},\ }\href {https://doi.org/10.1103/PhysRevA.59.1038} {\bibfield  {journal} {\bibinfo  {journal} {Phys. Rev. A}\ }\textbf {\bibinfo {volume} {59}},\ \bibinfo {pages} {1038} (\bibinfo {year} {1999})}\BibitemShut {NoStop}%
\bibitem [{\citenamefont {Xuereb}\ \emph {et~al.}(2011)\citenamefont {Xuereb}, \citenamefont {Schnabel},\ and\ \citenamefont {Hammerer}}]{Xuereb2011a}%
  \BibitemOpen
  \bibfield  {author} {\bibinfo {author} {\bibfnamefont {A.}~\bibnamefont {Xuereb}}, \bibinfo {author} {\bibfnamefont {R.}~\bibnamefont {Schnabel}},\ and\ \bibinfo {author} {\bibfnamefont {K.}~\bibnamefont {Hammerer}},\ }\bibfield  {title} {\bibinfo {title} {{Dissipative Optomechanics in a Michelson-Sagnac Interferometer}},\ }\href {https://doi.org/10.1103/PhysRevLett.107.213604} {\bibfield  {journal} {\bibinfo  {journal} {Phys. Rev. Lett.}\ }\textbf {\bibinfo {volume} {107}},\ \bibinfo {pages} {213604} (\bibinfo {year} {2011})}\BibitemShut {NoStop}%
\bibitem [{\citenamefont {Barzanjeh}\ \emph {et~al.}(2015)\citenamefont {Barzanjeh}, \citenamefont {Guha}, \citenamefont {Weedbrook}, \citenamefont {Vitali}, \citenamefont {Shapiro},\ and\ \citenamefont {Pirandola}}]{Barzanjeh2015microwave}%
  \BibitemOpen
  \bibfield  {author} {\bibinfo {author} {\bibfnamefont {S.}~\bibnamefont {Barzanjeh}}, \bibinfo {author} {\bibfnamefont {S.}~\bibnamefont {Guha}}, \bibinfo {author} {\bibfnamefont {C.}~\bibnamefont {Weedbrook}}, \bibinfo {author} {\bibfnamefont {D.}~\bibnamefont {Vitali}}, \bibinfo {author} {\bibfnamefont {J.~H.}\ \bibnamefont {Shapiro}},\ and\ \bibinfo {author} {\bibfnamefont {S.}~\bibnamefont {Pirandola}},\ }\bibfield  {title} {\bibinfo {title} {{Microwave quantum illumination}},\ }\href@noop {} {\bibfield  {journal} {\bibinfo  {journal} {Phys. Rev. Lett.}\ }\textbf {\bibinfo {volume} {114}},\ \bibinfo {pages} {080503} (\bibinfo {year} {2015})}\BibitemShut {NoStop}%
\bibitem [{\citenamefont {Nielsen}\ and\ \citenamefont {Chuang}(2000)}]{Nielsen2000a}%
  \BibitemOpen
  \bibfield  {author} {\bibinfo {author} {\bibfnamefont {M.~A.}\ \bibnamefont {Nielsen}}\ and\ \bibinfo {author} {\bibfnamefont {I.~L.}\ \bibnamefont {Chuang}},\ }\href@noop {} {\emph {\bibinfo {title} {{Quantum Computation and Quantum Information}}}}\ (\bibinfo  {publisher} {Cambridge University Press},\ \bibinfo {year} {2000})\BibitemShut {NoStop}%
\bibitem [{\citenamefont {Weedbrook}\ \emph {et~al.}(2012)\citenamefont {Weedbrook}, \citenamefont {Pirandola}, \citenamefont {Garc\'{\i}a-Patr\'on}, \citenamefont {Cerf}, \citenamefont {Ralph}, \citenamefont {Shapiro},\ and\ \citenamefont {Lloyd}}]{Weedbrook2012a}%
  \BibitemOpen
  \bibfield  {author} {\bibinfo {author} {\bibfnamefont {C.}~\bibnamefont {Weedbrook}}, \bibinfo {author} {\bibfnamefont {S.}~\bibnamefont {Pirandola}}, \bibinfo {author} {\bibfnamefont {R.}~\bibnamefont {Garc\'{\i}a-Patr\'on}}, \bibinfo {author} {\bibfnamefont {N.~J.}\ \bibnamefont {Cerf}}, \bibinfo {author} {\bibfnamefont {T.~C.}\ \bibnamefont {Ralph}}, \bibinfo {author} {\bibfnamefont {J.~H.}\ \bibnamefont {Shapiro}},\ and\ \bibinfo {author} {\bibfnamefont {S.}~\bibnamefont {Lloyd}},\ }\bibfield  {title} {\bibinfo {title} {{Gaussian quantum information}},\ }\href {https://doi.org/10.1103/RevModPhys.84.621} {\bibfield  {journal} {\bibinfo  {journal} {Rev. Mod. Phys.}\ }\textbf {\bibinfo {volume} {84}},\ \bibinfo {pages} {621} (\bibinfo {year} {2012})}\BibitemShut {NoStop}%
\bibitem [{\citenamefont {Braunstein}\ and\ \citenamefont {van Loock}(2005)}]{Braunstein2005a}%
  \BibitemOpen
  \bibfield  {author} {\bibinfo {author} {\bibfnamefont {S.~L.}\ \bibnamefont {Braunstein}}\ and\ \bibinfo {author} {\bibfnamefont {P.}~\bibnamefont {van Loock}},\ }\bibfield  {title} {\bibinfo {title} {{Quantum information with continuous variables}},\ }\href {https://doi.org/10.1103/RevModPhys.77.513} {\bibfield  {journal} {\bibinfo  {journal} {Rev. Mod. Phys.}\ }\textbf {\bibinfo {volume} {77}},\ \bibinfo {pages} {513} (\bibinfo {year} {2005})}\BibitemShut {NoStop}%
\bibitem [{\citenamefont {Kotler}\ \emph {et~al.}(2021)\citenamefont {Kotler}, \citenamefont {Peterson}, \citenamefont {Shojaee}, \citenamefont {Lecocq}, \citenamefont {Cicak}, \citenamefont {Kwiatkowski}, \citenamefont {Geller}, \citenamefont {Glancy}, \citenamefont {Knill}, \citenamefont {Simmonds} \emph {et~al.}}]{Kotler2021direct}%
  \BibitemOpen
  \bibfield  {author} {\bibinfo {author} {\bibfnamefont {S.}~\bibnamefont {Kotler}}, \bibinfo {author} {\bibfnamefont {G.~A.}\ \bibnamefont {Peterson}}, \bibinfo {author} {\bibfnamefont {E.}~\bibnamefont {Shojaee}}, \bibinfo {author} {\bibfnamefont {F.}~\bibnamefont {Lecocq}}, \bibinfo {author} {\bibfnamefont {K.}~\bibnamefont {Cicak}}, \bibinfo {author} {\bibfnamefont {A.}~\bibnamefont {Kwiatkowski}}, \bibinfo {author} {\bibfnamefont {S.}~\bibnamefont {Geller}}, \bibinfo {author} {\bibfnamefont {S.}~\bibnamefont {Glancy}}, \bibinfo {author} {\bibfnamefont {E.}~\bibnamefont {Knill}}, \bibinfo {author} {\bibfnamefont {R.~W.}\ \bibnamefont {Simmonds}}, \emph {et~al.},\ }\bibfield  {title} {\bibinfo {title} {{Direct observation of deterministic macroscopic entanglement}},\ }\href@noop {} {\bibfield  {journal} {\bibinfo  {journal} {Science}\ }\textbf {\bibinfo {volume} {372}},\ \bibinfo {pages} {622} (\bibinfo {year} {2021})}\BibitemShut {NoStop}%
\bibitem [{\citenamefont {Cosco}\ \emph {et~al.}(2021)\citenamefont {Cosco}, \citenamefont {Pedernales},\ and\ \citenamefont {Plenio}}]{Cosco2021a}%
  \BibitemOpen
  \bibfield  {author} {\bibinfo {author} {\bibfnamefont {F.}~\bibnamefont {Cosco}}, \bibinfo {author} {\bibfnamefont {J.~S.}\ \bibnamefont {Pedernales}},\ and\ \bibinfo {author} {\bibfnamefont {M.~B.}\ \bibnamefont {Plenio}},\ }\bibfield  {title} {\bibinfo {title} {{Enhanced force sensitivity and entanglement in periodically driven optomechanics}},\ }\href {https://doi.org/10.1103/PhysRevA.103.L061501} {\bibfield  {journal} {\bibinfo  {journal} {Phys. Rev. A}\ }\textbf {\bibinfo {volume} {103}},\ \bibinfo {pages} {L061501} (\bibinfo {year} {2021})}\BibitemShut {NoStop}%
\bibitem [{\citenamefont {Weiss}\ \emph {et~al.}(2021)\citenamefont {Weiss}, \citenamefont {Roda-Llordes}, \citenamefont {Torrontegui}, \citenamefont {Aspelmeyer},\ and\ \citenamefont {Romero-Isart}}]{Weiss2021a}%
  \BibitemOpen
  \bibfield  {author} {\bibinfo {author} {\bibfnamefont {T.}~\bibnamefont {Weiss}}, \bibinfo {author} {\bibfnamefont {M.}~\bibnamefont {Roda-Llordes}}, \bibinfo {author} {\bibfnamefont {E.}~\bibnamefont {Torrontegui}}, \bibinfo {author} {\bibfnamefont {M.}~\bibnamefont {Aspelmeyer}},\ and\ \bibinfo {author} {\bibfnamefont {O.}~\bibnamefont {Romero-Isart}},\ }\bibfield  {title} {\bibinfo {title} {{Large Quantum Delocalization of a Levitated Nanoparticle Using Optimal Control: Applications for Force Sensing and Entangling via Weak Forces}},\ }\href {https://doi.org/10.1103/PhysRevLett.127.023601} {\bibfield  {journal} {\bibinfo  {journal} {Phys. Rev. Lett.}\ }\textbf {\bibinfo {volume} {127}},\ \bibinfo {pages} {023601} (\bibinfo {year} {2021})}\BibitemShut {NoStop}%
\bibitem [{\citenamefont {Cong}(2014)}]{cong2014control}%
  \BibitemOpen
  \bibfield  {author} {\bibinfo {author} {\bibfnamefont {S.}~\bibnamefont {Cong}},\ }\href@noop {} {\emph {\bibinfo {title} {{Control of quantum systems: theory and methods}}}}\ (\bibinfo  {publisher} {John Wiley \& Sons},\ \bibinfo {year} {2014})\BibitemShut {NoStop}%
\bibitem [{\citenamefont {Glaser}\ \emph {et~al.}(2015)\citenamefont {Glaser}, \citenamefont {Boscain}, \citenamefont {Calarco}, \citenamefont {Koch}, \citenamefont {K{\"o}ckenberger}, \citenamefont {Kosloff}, \citenamefont {Kuprov}, \citenamefont {Luy}, \citenamefont {Schirmer}, \citenamefont {Schulte-Herbr{\"u}ggen} \emph {et~al.}}]{glaser2015training}%
  \BibitemOpen
  \bibfield  {author} {\bibinfo {author} {\bibfnamefont {S.~J.}\ \bibnamefont {Glaser}}, \bibinfo {author} {\bibfnamefont {U.}~\bibnamefont {Boscain}}, \bibinfo {author} {\bibfnamefont {T.}~\bibnamefont {Calarco}}, \bibinfo {author} {\bibfnamefont {C.~P.}\ \bibnamefont {Koch}}, \bibinfo {author} {\bibfnamefont {W.}~\bibnamefont {K{\"o}ckenberger}}, \bibinfo {author} {\bibfnamefont {R.}~\bibnamefont {Kosloff}}, \bibinfo {author} {\bibfnamefont {I.}~\bibnamefont {Kuprov}}, \bibinfo {author} {\bibfnamefont {B.}~\bibnamefont {Luy}}, \bibinfo {author} {\bibfnamefont {S.}~\bibnamefont {Schirmer}}, \bibinfo {author} {\bibfnamefont {T.}~\bibnamefont {Schulte-Herbr{\"u}ggen}}, \emph {et~al.},\ }\bibfield  {title} {\bibinfo {title} {{Training Schrodinger's cat: Quantum optimal control: Strategic report on current status, visions and goals for research in Europe}},\ }\href@noop {} {\bibfield  {journal} {\bibinfo  {journal} {EPJ D}\ }\textbf {\bibinfo {volume} {69}},\ \bibinfo {pages} {1} (\bibinfo {year} {2015})}\BibitemShut {NoStop}%
\bibitem [{\citenamefont {Koch}(2016)}]{koch2016controlling}%
  \BibitemOpen
  \bibfield  {author} {\bibinfo {author} {\bibfnamefont {C.~P.}\ \bibnamefont {Koch}},\ }\bibfield  {title} {\bibinfo {title} {{Controlling open quantum systems: tools, achievements, and limitations}},\ }\href@noop {} {\bibfield  {journal} {\bibinfo  {journal} {Journal of Physics: Condensed Matter}\ }\textbf {\bibinfo {volume} {28}},\ \bibinfo {pages} {213001} (\bibinfo {year} {2016})}\BibitemShut {NoStop}%
\bibitem [{\citenamefont {Borz{\`\i}}\ \emph {et~al.}(2017)\citenamefont {Borz{\`\i}}, \citenamefont {Ciaramella},\ and\ \citenamefont {Sprengel}}]{borzi2017formulation}%
  \BibitemOpen
  \bibfield  {author} {\bibinfo {author} {\bibfnamefont {A.}~\bibnamefont {Borz{\`\i}}}, \bibinfo {author} {\bibfnamefont {G.}~\bibnamefont {Ciaramella}},\ and\ \bibinfo {author} {\bibfnamefont {M.}~\bibnamefont {Sprengel}},\ }\href@noop {} {\emph {\bibinfo {title} {{Formulation and numerical solution of quantum control problems}}}}\ (\bibinfo  {publisher} {SIAM},\ \bibinfo {year} {2017})\BibitemShut {NoStop}%
\bibitem [{\citenamefont {d'Alessandro}(2021)}]{d2021introduction}%
  \BibitemOpen
  \bibfield  {author} {\bibinfo {author} {\bibfnamefont {D.}~\bibnamefont {d'Alessandro}},\ }\href@noop {} {\emph {\bibinfo {title} {{Introduction to quantum control and dynamics}}}}\ (\bibinfo  {publisher} {CRC press},\ \bibinfo {year} {2021})\BibitemShut {NoStop}%
\bibitem [{\citenamefont {Vitanov}\ \emph {et~al.}(2017)\citenamefont {Vitanov}, \citenamefont {Rangelov}, \citenamefont {Shore},\ and\ \citenamefont {Bergmann}}]{Vitanov2017a}%
  \BibitemOpen
  \bibfield  {author} {\bibinfo {author} {\bibfnamefont {N.~V.}\ \bibnamefont {Vitanov}}, \bibinfo {author} {\bibfnamefont {A.~A.}\ \bibnamefont {Rangelov}}, \bibinfo {author} {\bibfnamefont {B.~W.}\ \bibnamefont {Shore}},\ and\ \bibinfo {author} {\bibfnamefont {K.}~\bibnamefont {Bergmann}},\ }\bibfield  {title} {\bibinfo {title} {{Stimulated Raman adiabatic passage in physics, chemistry, and beyond}},\ }\href {https://doi.org/10.1103/RevModPhys.89.015006} {\bibfield  {journal} {\bibinfo  {journal} {Rev. Mod. Phys.}\ }\textbf {\bibinfo {volume} {89}},\ \bibinfo {pages} {015006} (\bibinfo {year} {2017})}\BibitemShut {NoStop}%
\bibitem [{\citenamefont {Doria}\ \emph {et~al.}(2011)\citenamefont {Doria}, \citenamefont {Calarco},\ and\ \citenamefont {Montangero}}]{Doria2011a}%
  \BibitemOpen
  \bibfield  {author} {\bibinfo {author} {\bibfnamefont {P.}~\bibnamefont {Doria}}, \bibinfo {author} {\bibfnamefont {T.}~\bibnamefont {Calarco}},\ and\ \bibinfo {author} {\bibfnamefont {S.}~\bibnamefont {Montangero}},\ }\bibfield  {title} {\bibinfo {title} {{Optimal Control Technique for Many-Body Quantum Dynamics}},\ }\href {https://doi.org/10.1103/PhysRevLett.106.190501} {\bibfield  {journal} {\bibinfo  {journal} {Phys. Rev. Lett.}\ }\textbf {\bibinfo {volume} {106}},\ \bibinfo {pages} {190501} (\bibinfo {year} {2011})}\BibitemShut {NoStop}%
\bibitem [{\citenamefont {Muller}\ \emph {et~al.}(2022)\citenamefont {Muller}, \citenamefont {Said}, \citenamefont {Jelezko}, \citenamefont {Calarco},\ and\ \citenamefont {Montangero}}]{Muller2022a}%
  \BibitemOpen
  \bibfield  {author} {\bibinfo {author} {\bibfnamefont {M.~M.}\ \bibnamefont {Muller}}, \bibinfo {author} {\bibfnamefont {R.~S.}\ \bibnamefont {Said}}, \bibinfo {author} {\bibfnamefont {F.}~\bibnamefont {Jelezko}}, \bibinfo {author} {\bibfnamefont {T.}~\bibnamefont {Calarco}},\ and\ \bibinfo {author} {\bibfnamefont {S.}~\bibnamefont {Montangero}},\ }\bibfield  {title} {\bibinfo {title} {{One decade of quantum optimal control in the chopped random basis}},\ }\href {https://doi.org/10.1088/1361-6633/ac723c} {\bibfield  {journal} {\bibinfo  {journal} {Reports on Progress in Physics}\ }\textbf {\bibinfo {volume} {85}},\ \bibinfo {pages} {076001} (\bibinfo {year} {2022})}\BibitemShut {NoStop}%
\bibitem [{\citenamefont {Khaneja}\ \emph {et~al.}(2005)\citenamefont {Khaneja}, \citenamefont {Reiss}, \citenamefont {Kehlet}, \citenamefont {Schulte-Herbruggen},\ and\ \citenamefont {Glaser}}]{Khaneja2005a}%
  \BibitemOpen
  \bibfield  {author} {\bibinfo {author} {\bibfnamefont {N.}~\bibnamefont {Khaneja}}, \bibinfo {author} {\bibfnamefont {T.}~\bibnamefont {Reiss}}, \bibinfo {author} {\bibfnamefont {C.}~\bibnamefont {Kehlet}}, \bibinfo {author} {\bibfnamefont {T.}~\bibnamefont {Schulte-Herbruggen}},\ and\ \bibinfo {author} {\bibfnamefont {S.~J.}\ \bibnamefont {Glaser}},\ }\bibfield  {title} {\bibinfo {title} {{Optimal control of coupled spin dynamics: design of NMR pulse sequences by gradient ascent algorithms}},\ }\href {https://doi.org/10.1016/j.jmr.2004.11.004} {\bibfield  {journal} {\bibinfo  {journal} {Journal of Magnetic Resonance}\ }\textbf {\bibinfo {volume} {172}},\ \bibinfo {pages} {296} (\bibinfo {year} {2005})}\BibitemShut {NoStop}%
\bibitem [{\citenamefont {Goerz}\ \emph {et~al.}(2022)\citenamefont {Goerz}, \citenamefont {Carrasco},\ and\ \citenamefont {Malinovsky}}]{Goerz2022a}%
  \BibitemOpen
  \bibfield  {author} {\bibinfo {author} {\bibfnamefont {M.~H.}\ \bibnamefont {Goerz}}, \bibinfo {author} {\bibfnamefont {S.~C.}\ \bibnamefont {Carrasco}},\ and\ \bibinfo {author} {\bibfnamefont {V.~S.}\ \bibnamefont {Malinovsky}},\ }\bibfield  {title} {\bibinfo {title} {{Quantum Optimal Control via Semi-Automatic Differentiation}},\ }\href {https://doi.org/10.22331/q-2022-12-07-871} {\bibfield  {journal} {\bibinfo  {journal} {Quantum}\ }\textbf {\bibinfo {volume} {6}},\ \bibinfo {pages} {871} (\bibinfo {year} {2022})}\BibitemShut {NoStop}%
\bibitem [{\citenamefont {Niu}\ \emph {et~al.}(2019)\citenamefont {Niu}, \citenamefont {Boixo}, \citenamefont {Smelyanskiy},\ and\ \citenamefont {Neven}}]{Niu2019a}%
  \BibitemOpen
  \bibfield  {author} {\bibinfo {author} {\bibfnamefont {M.~Y.}\ \bibnamefont {Niu}}, \bibinfo {author} {\bibfnamefont {S.}~\bibnamefont {Boixo}}, \bibinfo {author} {\bibfnamefont {V.~N.}\ \bibnamefont {Smelyanskiy}},\ and\ \bibinfo {author} {\bibfnamefont {H.}~\bibnamefont {Neven}},\ }\bibfield  {title} {\bibinfo {title} {{Universal quantum control through deep reinforcement learning}},\ }\href {https://doi.org/10.1038/s41534-019-0141-3} {\bibfield  {journal} {\bibinfo  {journal} {npj Quantum Information}\ }\textbf {\bibinfo {volume} {5}},\ \bibinfo {pages} {33} (\bibinfo {year} {2019})}\BibitemShut {NoStop}%
\bibitem [{\citenamefont {Sivak}\ \emph {et~al.}(2022)\citenamefont {Sivak}, \citenamefont {Eickbusch}, \citenamefont {Liu}, \citenamefont {Royer}, \citenamefont {Tsioutsios},\ and\ \citenamefont {Devoret}}]{Sivak2022a}%
  \BibitemOpen
  \bibfield  {author} {\bibinfo {author} {\bibfnamefont {V.~V.}\ \bibnamefont {Sivak}}, \bibinfo {author} {\bibfnamefont {A.}~\bibnamefont {Eickbusch}}, \bibinfo {author} {\bibfnamefont {H.}~\bibnamefont {Liu}}, \bibinfo {author} {\bibfnamefont {B.}~\bibnamefont {Royer}}, \bibinfo {author} {\bibfnamefont {I.}~\bibnamefont {Tsioutsios}},\ and\ \bibinfo {author} {\bibfnamefont {M.~H.}\ \bibnamefont {Devoret}},\ }\bibfield  {title} {\bibinfo {title} {{Model-Free Quantum Control with Reinforcement Learning}},\ }\href {https://doi.org/10.1103/PhysRevX.12.011059} {\bibfield  {journal} {\bibinfo  {journal} {Phys. Rev. X}\ }\textbf {\bibinfo {volume} {12}},\ \bibinfo {pages} {011059} (\bibinfo {year} {2022})}\BibitemShut {NoStop}%
\bibitem [{\citenamefont {Konnov}\ and\ \citenamefont {Krotov}(1999)}]{Konnov1999b}%
  \BibitemOpen
  \bibfield  {author} {\bibinfo {author} {\bibfnamefont {A.}~\bibnamefont {Konnov}}\ and\ \bibinfo {author} {\bibfnamefont {V.~F.}\ \bibnamefont {Krotov}},\ }\bibfield  {title} {\bibinfo {title} {{On global methods for the successive improvement of control processes}},\ }\href@noop {} {\bibfield  {journal} {\bibinfo  {journal} {Avtomatika i Telemekhanika}\ ,\ \bibinfo {pages} {77}} (\bibinfo {year} {1999})}\BibitemShut {NoStop}%
\bibitem [{\citenamefont {Reich}\ \emph {et~al.}(2012)\citenamefont {Reich}, \citenamefont {Ndong},\ and\ \citenamefont {Koch}}]{Reich2012v}%
  \BibitemOpen
  \bibfield  {author} {\bibinfo {author} {\bibfnamefont {D.~M.}\ \bibnamefont {Reich}}, \bibinfo {author} {\bibfnamefont {M.}~\bibnamefont {Ndong}},\ and\ \bibinfo {author} {\bibfnamefont {C.~P.}\ \bibnamefont {Koch}},\ }\bibfield  {title} {\bibinfo {title} {{Monotonically convergent optimization in quantum control using Krotov's method}},\ }\href {https://doi.org/10.1063/1.3691827} {\bibfield  {journal} {\bibinfo  {journal} {J. Chem. Phys.}\ }\textbf {\bibinfo {volume} {136}},\ \bibinfo {pages} {104103} (\bibinfo {year} {2012})}\BibitemShut {NoStop}%
\bibitem [{\citenamefont {Tannor}\ \emph {et~al.}(1992)\citenamefont {Tannor}, \citenamefont {Kazakov},\ and\ \citenamefont {Orlov}}]{Tannor1992h}%
  \BibitemOpen
  \bibfield  {author} {\bibinfo {author} {\bibfnamefont {D.~J.}\ \bibnamefont {Tannor}}, \bibinfo {author} {\bibfnamefont {V.}~\bibnamefont {Kazakov}},\ and\ \bibinfo {author} {\bibfnamefont {V.}~\bibnamefont {Orlov}},\ }\bibfield  {title} {\bibinfo {title} {{Control of Photochemical Branching: Novel Procedures for Finding Optimal Pulses and Global Upper Bounds}},\ }\bibfield  {booktitle} {\emph {\bibinfo {booktitle} {{Time-Dependent Quantum Molecular Dynamics}}},\ }\href {https://doi.org/10.1007/978-1-4899-2326-4_24} {\bibfield  {journal} {\bibinfo  {journal} {Time-Dependent Quantum Molecular Dynamics}\ ,\ \bibinfo {pages} {347}} (\bibinfo {year} {1992})}\BibitemShut {NoStop}%
\bibitem [{\citenamefont {J\"ager}\ \emph {et~al.}(2014)\citenamefont {J\"ager}, \citenamefont {Reich}, \citenamefont {Goerz}, \citenamefont {Koch},\ and\ \citenamefont {Hohenester}}]{Jager2014a}%
  \BibitemOpen
  \bibfield  {author} {\bibinfo {author} {\bibfnamefont {G.}~\bibnamefont {J\"ager}}, \bibinfo {author} {\bibfnamefont {D.~M.}\ \bibnamefont {Reich}}, \bibinfo {author} {\bibfnamefont {M.~H.}\ \bibnamefont {Goerz}}, \bibinfo {author} {\bibfnamefont {C.~P.}\ \bibnamefont {Koch}},\ and\ \bibinfo {author} {\bibfnamefont {U.}~\bibnamefont {Hohenester}},\ }\bibfield  {title} {\bibinfo {title} {{Optimal quantum control of Bose-Einstein condensates in magnetic microtraps: Comparison of gradient-ascent-pulse-engineering and Krotov optimization schemes}},\ }\href {https://doi.org/10.1103/PhysRevA.90.033628} {\bibfield  {journal} {\bibinfo  {journal} {Phys. Rev. A}\ }\textbf {\bibinfo {volume} {90}},\ \bibinfo {pages} {033628} (\bibinfo {year} {2014})}\BibitemShut {NoStop}%
\bibitem [{\citenamefont {Goerz}\ \emph {et~al.}(2019)\citenamefont {Goerz}, \citenamefont {Basilewitsch}, \citenamefont {Gago-Encinas}, \citenamefont {Krauss}, \citenamefont {Horn}, \citenamefont {Reich},\ and\ \citenamefont {Koch}}]{Goerz2019w}%
  \BibitemOpen
  \bibfield  {author} {\bibinfo {author} {\bibfnamefont {M.~H.}\ \bibnamefont {Goerz}}, \bibinfo {author} {\bibfnamefont {D.}~\bibnamefont {Basilewitsch}}, \bibinfo {author} {\bibfnamefont {F.}~\bibnamefont {Gago-Encinas}}, \bibinfo {author} {\bibfnamefont {M.~G.}\ \bibnamefont {Krauss}}, \bibinfo {author} {\bibfnamefont {K.~P.}\ \bibnamefont {Horn}}, \bibinfo {author} {\bibfnamefont {D.~M.}\ \bibnamefont {Reich}},\ and\ \bibinfo {author} {\bibfnamefont {C.~P.}\ \bibnamefont {Koch}},\ }\bibfield  {title} {\bibinfo {title} {{{Krotov: A Python implementation of Krotov's method for quantum optimal control}}},\ }\href {https://doi.org/10.21468/SciPostPhys.7.6.080} {\bibfield  {journal} {\bibinfo  {journal} {SciPost Phys.}\ }\textbf {\bibinfo {volume} {7}},\ \bibinfo {pages} {80} (\bibinfo {year} {2019})}\BibitemShut {NoStop}%
\bibitem [{\citenamefont {Sklarz}\ and\ \citenamefont {Tannor}(2002)}]{sklarz2002loading}%
  \BibitemOpen
  \bibfield  {author} {\bibinfo {author} {\bibfnamefont {S.~E.}\ \bibnamefont {Sklarz}}\ and\ \bibinfo {author} {\bibfnamefont {D.~J.}\ \bibnamefont {Tannor}},\ }\bibfield  {title} {\bibinfo {title} {{Loading a Bose-Einstein condensate onto an optical lattice: An application of optimal control theory to the nonlinear Schr{\"o}dinger equation}},\ }\href@noop {} {\bibfield  {journal} {\bibinfo  {journal} {Phys. Rev. A}\ }\textbf {\bibinfo {volume} {66}},\ \bibinfo {pages} {053619} (\bibinfo {year} {2002})}\BibitemShut {NoStop}%
\bibitem [{\citenamefont {Somloi}\ \emph {et~al.}(1993)\citenamefont {Somloi}, \citenamefont {Kazakov},\ and\ \citenamefont {Tannor}}]{somloi1993controlled}%
  \BibitemOpen
  \bibfield  {author} {\bibinfo {author} {\bibfnamefont {J.}~\bibnamefont {Somloi}}, \bibinfo {author} {\bibfnamefont {V.~A.}\ \bibnamefont {Kazakov}},\ and\ \bibinfo {author} {\bibfnamefont {D.~J.}\ \bibnamefont {Tannor}},\ }\bibfield  {title} {\bibinfo {title} {{Controlled dissociation of I2 via optical transitions between the X and B electronic states}},\ }\href@noop {} {\bibfield  {journal} {\bibinfo  {journal} {Chem. Phys.}\ }\textbf {\bibinfo {volume} {172}},\ \bibinfo {pages} {85} (\bibinfo {year} {1993})}\BibitemShut {NoStop}%
\bibitem [{\citenamefont {Hwang}\ and\ \citenamefont {Goan}(2012)}]{hwang2012optimal}%
  \BibitemOpen
  \bibfield  {author} {\bibinfo {author} {\bibfnamefont {B.}~\bibnamefont {Hwang}}\ and\ \bibinfo {author} {\bibfnamefont {H.-S.}\ \bibnamefont {Goan}},\ }\bibfield  {title} {\bibinfo {title} {{Optimal control for non-Markovian open quantum systems}},\ }\href@noop {} {\bibfield  {journal} {\bibinfo  {journal} {Phys. Rev. A}\ }\textbf {\bibinfo {volume} {85}},\ \bibinfo {pages} {032321} (\bibinfo {year} {2012})}\BibitemShut {NoStop}%
\bibitem [{\citenamefont {Watts}\ \emph {et~al.}(2015)\citenamefont {Watts}, \citenamefont {Vala}, \citenamefont {M\"uller}, \citenamefont {Calarco}, \citenamefont {Whaley}, \citenamefont {Reich}, \citenamefont {Goerz},\ and\ \citenamefont {Koch}}]{Watts2015a}%
  \BibitemOpen
  \bibfield  {author} {\bibinfo {author} {\bibfnamefont {P.}~\bibnamefont {Watts}}, \bibinfo {author} {\bibfnamefont {J.~c.~v.}\ \bibnamefont {Vala}}, \bibinfo {author} {\bibfnamefont {M.~M.}\ \bibnamefont {M\"uller}}, \bibinfo {author} {\bibfnamefont {T.}~\bibnamefont {Calarco}}, \bibinfo {author} {\bibfnamefont {K.~B.}\ \bibnamefont {Whaley}}, \bibinfo {author} {\bibfnamefont {D.~M.}\ \bibnamefont {Reich}}, \bibinfo {author} {\bibfnamefont {M.~H.}\ \bibnamefont {Goerz}},\ and\ \bibinfo {author} {\bibfnamefont {C.~P.}\ \bibnamefont {Koch}},\ }\bibfield  {title} {\bibinfo {title} {{Optimizing for an arbitrary perfect entangler. I. Functionals}},\ }\href {https://doi.org/10.1103/PhysRevA.91.062306} {\bibfield  {journal} {\bibinfo  {journal} {Phys. Rev. A}\ }\textbf {\bibinfo {volume} {91}},\ \bibinfo {pages} {062306} (\bibinfo {year} {2015})}\BibitemShut {NoStop}%
\bibitem [{\citenamefont {Goerz}\ \emph {et~al.}(2015)\citenamefont {Goerz}, \citenamefont {Gualdi}, \citenamefont {Reich}, \citenamefont {Koch}, \citenamefont {Motzoi}, \citenamefont {Whaley}, \citenamefont {Vala}, \citenamefont {M\"uller}, \citenamefont {Montangero},\ and\ \citenamefont {Calarco}}]{Goerz2015a}%
  \BibitemOpen
  \bibfield  {author} {\bibinfo {author} {\bibfnamefont {M.~H.}\ \bibnamefont {Goerz}}, \bibinfo {author} {\bibfnamefont {G.}~\bibnamefont {Gualdi}}, \bibinfo {author} {\bibfnamefont {D.~M.}\ \bibnamefont {Reich}}, \bibinfo {author} {\bibfnamefont {C.~P.}\ \bibnamefont {Koch}}, \bibinfo {author} {\bibfnamefont {F.}~\bibnamefont {Motzoi}}, \bibinfo {author} {\bibfnamefont {K.~B.}\ \bibnamefont {Whaley}}, \bibinfo {author} {\bibfnamefont {J.~c.~v.}\ \bibnamefont {Vala}}, \bibinfo {author} {\bibfnamefont {M.~M.}\ \bibnamefont {M\"uller}}, \bibinfo {author} {\bibfnamefont {S.}~\bibnamefont {Montangero}},\ and\ \bibinfo {author} {\bibfnamefont {T.}~\bibnamefont {Calarco}},\ }\bibfield  {title} {\bibinfo {title} {{Optimizing for an arbitrary perfect entangler. II. Application}},\ }\href {https://doi.org/10.1103/PhysRevA.91.062307} {\bibfield  {journal} {\bibinfo  {journal} {Phys. Rev. A}\ }\textbf {\bibinfo {volume} {91}},\ \bibinfo {pages} {062307} (\bibinfo {year} {2015})}\BibitemShut {NoStop}%
\bibitem [{\citenamefont {Rossi}\ \emph {et~al.}(2018)\citenamefont {Rossi}, \citenamefont {Mason}, \citenamefont {Chen}, \citenamefont {Tsaturyan},\ and\ \citenamefont {Schliesser}}]{Rossi2018a}%
  \BibitemOpen
  \bibfield  {author} {\bibinfo {author} {\bibfnamefont {M.}~\bibnamefont {Rossi}}, \bibinfo {author} {\bibfnamefont {D.}~\bibnamefont {Mason}}, \bibinfo {author} {\bibfnamefont {J.}~\bibnamefont {Chen}}, \bibinfo {author} {\bibfnamefont {Y.}~\bibnamefont {Tsaturyan}},\ and\ \bibinfo {author} {\bibfnamefont {A.}~\bibnamefont {Schliesser}},\ }\bibfield  {title} {\bibinfo {title} {{Measurement-based quantum control of mechanical motion}},\ }\href {https://doi.org/10.1038/s41586-018-0643-8} {\bibfield  {journal} {\bibinfo  {journal} {Nature}\ }\textbf {\bibinfo {volume} {563}},\ \bibinfo {pages} {53} (\bibinfo {year} {2018})}\BibitemShut {NoStop}%
\bibitem [{\citenamefont {Miki}\ \emph {et~al.}(2023)\citenamefont {Miki}, \citenamefont {Matsumoto}, \citenamefont {Matsumura}, \citenamefont {Shichijo}, \citenamefont {Sugiyama}, \citenamefont {Yamamoto},\ and\ \citenamefont {Yamamoto}}]{miki2023generating}%
  \BibitemOpen
  \bibfield  {author} {\bibinfo {author} {\bibfnamefont {D.}~\bibnamefont {Miki}}, \bibinfo {author} {\bibfnamefont {N.}~\bibnamefont {Matsumoto}}, \bibinfo {author} {\bibfnamefont {A.}~\bibnamefont {Matsumura}}, \bibinfo {author} {\bibfnamefont {T.}~\bibnamefont {Shichijo}}, \bibinfo {author} {\bibfnamefont {Y.}~\bibnamefont {Sugiyama}}, \bibinfo {author} {\bibfnamefont {K.}~\bibnamefont {Yamamoto}},\ and\ \bibinfo {author} {\bibfnamefont {N.}~\bibnamefont {Yamamoto}},\ }\bibfield  {title} {\bibinfo {title} {{Generating quantum entanglement between macroscopic objects with continuous measurement and feedback control}},\ }\href@noop {} {\bibfield  {journal} {\bibinfo  {journal} {Phys. Rev. A}\ }\textbf {\bibinfo {volume} {107}},\ \bibinfo {pages} {032410} (\bibinfo {year} {2023})}\BibitemShut {NoStop}%
\bibitem [{\citenamefont {Wang}\ and\ \citenamefont {Clerk}(2013)}]{Wang2013a}%
  \BibitemOpen
  \bibfield  {author} {\bibinfo {author} {\bibfnamefont {Y.-D.}\ \bibnamefont {Wang}}\ and\ \bibinfo {author} {\bibfnamefont {A.~A.}\ \bibnamefont {Clerk}},\ }\bibfield  {title} {\bibinfo {title} {{Reservoir-Engineered Entanglement in Optomechanical Systems}},\ }\href {https://doi.org/10.1103/PhysRevLett.110.253601} {\bibfield  {journal} {\bibinfo  {journal} {Phys. Rev. Lett.}\ }\textbf {\bibinfo {volume} {110}},\ \bibinfo {pages} {253601} (\bibinfo {year} {2013})}\BibitemShut {NoStop}%
\bibitem [{\citenamefont {Clarke}\ \emph {et~al.}(2020)\citenamefont {Clarke}, \citenamefont {Sahium}, \citenamefont {Khosla}, \citenamefont {Pikovski}, \citenamefont {Kim},\ and\ \citenamefont {Vanner}}]{Clarke2020a}%
  \BibitemOpen
  \bibfield  {author} {\bibinfo {author} {\bibfnamefont {J.}~\bibnamefont {Clarke}}, \bibinfo {author} {\bibfnamefont {P.}~\bibnamefont {Sahium}}, \bibinfo {author} {\bibfnamefont {K.~E.}\ \bibnamefont {Khosla}}, \bibinfo {author} {\bibfnamefont {I.}~\bibnamefont {Pikovski}}, \bibinfo {author} {\bibfnamefont {M.~S.}\ \bibnamefont {Kim}},\ and\ \bibinfo {author} {\bibfnamefont {M.~R.}\ \bibnamefont {Vanner}},\ }\bibfield  {title} {\bibinfo {title} {{Generating mechanical and optomechanical entanglement via pulsed interaction and measurement}},\ }\href {https://doi.org/10.1088/1367-2630/ab7ddd} {\bibfield  {journal} {\bibinfo  {journal} {New Journal of Physics}\ }\textbf {\bibinfo {volume} {22}},\ \bibinfo {pages} {063001} (\bibinfo {year} {2020})}\BibitemShut {NoStop}%
\bibitem [{\citenamefont {Chen}\ \emph {et~al.}(2017)\citenamefont {Chen}, \citenamefont {Liao},\ and\ \citenamefont {Lin}}]{Chen2017a}%
  \BibitemOpen
  \bibfield  {author} {\bibinfo {author} {\bibfnamefont {R.-X.}\ \bibnamefont {Chen}}, \bibinfo {author} {\bibfnamefont {C.-G.}\ \bibnamefont {Liao}},\ and\ \bibinfo {author} {\bibfnamefont {X.-M.}\ \bibnamefont {Lin}},\ }\bibfield  {title} {\bibinfo {title} {{Dissipative generation of significant amount of mechanical entanglement in a coupled optomechanical system}},\ }\href {https://doi.org/10.1038/s41598-017-15032-1} {\bibfield  {journal} {\bibinfo  {journal} {Scientific Reports}\ }\textbf {\bibinfo {volume} {7}},\ \bibinfo {pages} {14497} (\bibinfo {year} {2017})}\BibitemShut {NoStop}%
\bibitem [{\citenamefont {Kuzyk}\ \emph {et~al.}(2013)\citenamefont {Kuzyk}, \citenamefont {van Enk},\ and\ \citenamefont {Wang}}]{Kuzyk2013a}%
  \BibitemOpen
  \bibfield  {author} {\bibinfo {author} {\bibfnamefont {M.~C.}\ \bibnamefont {Kuzyk}}, \bibinfo {author} {\bibfnamefont {S.~J.}\ \bibnamefont {van Enk}},\ and\ \bibinfo {author} {\bibfnamefont {H.}~\bibnamefont {Wang}},\ }\bibfield  {title} {\bibinfo {title} {{Generating robust optical entanglement in weak-coupling optomechanical systems}},\ }\href {https://doi.org/10.1103/PhysRevA.88.062341} {\bibfield  {journal} {\bibinfo  {journal} {Phys. Rev. A}\ }\textbf {\bibinfo {volume} {88}},\ \bibinfo {pages} {062341} (\bibinfo {year} {2013})}\BibitemShut {NoStop}%
\bibitem [{\citenamefont {Serafini}(2017)}]{Serafini2017a}%
  \BibitemOpen
  \bibfield  {author} {\bibinfo {author} {\bibfnamefont {A.}~\bibnamefont {Serafini}},\ }\href {https://doi.org/10.1201/9781315118727} {\bibinfo {title} {{Quantum Continuous Variables: A Primer of Theoretical Methods}}} (\bibinfo {year} {2017})\BibitemShut {NoStop}%
\bibitem [{\citenamefont {Agarwal}(2012)}]{Agarwal2012a}%
  \BibitemOpen
  \bibfield  {author} {\bibinfo {author} {\bibfnamefont {G.~S.}\ \bibnamefont {Agarwal}},\ }\href@noop {} {\emph {\bibinfo {title} {{Quantum optics}}}}\ (\bibinfo  {publisher} {Cambridge University Press},\ \bibinfo {year} {2012})\BibitemShut {NoStop}%
\bibitem [{\citenamefont {Castro}\ \emph {et~al.}(2019)\citenamefont {Castro}, \citenamefont {Appel},\ and\ \citenamefont {Rubio}}]{Castro2019a}%
  \BibitemOpen
  \bibfield  {author} {\bibinfo {author} {\bibfnamefont {A.}~\bibnamefont {Castro}}, \bibinfo {author} {\bibfnamefont {H.}~\bibnamefont {Appel}},\ and\ \bibinfo {author} {\bibfnamefont {A.}~\bibnamefont {Rubio}},\ }\bibfield  {title} {\bibinfo {title} {{Optimal control theory for quantum electrodynamics: an initial state problem}},\ }\href {https://doi.org/10.1140/epjb/e2019-100263-2} {\bibfield  {journal} {\bibinfo  {journal} {EPJ B}\ }\textbf {\bibinfo {volume} {92}},\ \bibinfo {pages} {223} (\bibinfo {year} {2019})}\BibitemShut {NoStop}%
\bibitem [{\citenamefont {Porotti}\ \emph {et~al.}(2022)\citenamefont {Porotti}, \citenamefont {Essig}, \citenamefont {Huard},\ and\ \citenamefont {Marquardt}}]{Porotti2022a}%
  \BibitemOpen
  \bibfield  {author} {\bibinfo {author} {\bibfnamefont {R.}~\bibnamefont {Porotti}}, \bibinfo {author} {\bibfnamefont {A.}~\bibnamefont {Essig}}, \bibinfo {author} {\bibfnamefont {B.}~\bibnamefont {Huard}},\ and\ \bibinfo {author} {\bibfnamefont {F.}~\bibnamefont {Marquardt}},\ }\bibfield  {title} {\bibinfo {title} {{Deep Reinforcement Learning for Quantum State Preparation with Weak Nonlinear Measurements}},\ }\href {https://doi.org/10.22331/q-2022-06-28-747} {\bibfield  {journal} {\bibinfo  {journal} {Quantum}\ }\textbf {\bibinfo {volume} {6}},\ \bibinfo {pages} {747} (\bibinfo {year} {2022})}\BibitemShut {NoStop}%
\bibitem [{\citenamefont {Kudra}\ \emph {et~al.}(2022)\citenamefont {Kudra}, \citenamefont {Kervinen}, \citenamefont {Strandberg}, \citenamefont {Ahmed}, \citenamefont {Scigliuzzo}, \citenamefont {Osman}, \citenamefont {Lozano}, \citenamefont {Thol\'en}, \citenamefont {Borgani}, \citenamefont {Haviland}, \citenamefont {Ferrini}, \citenamefont {Bylander}, \citenamefont {Kockum}, \citenamefont {Quijandr\'{\i}a}, \citenamefont {Delsing},\ and\ \citenamefont {Gasparinetti}}]{Kudra2022a}%
  \BibitemOpen
  \bibfield  {author} {\bibinfo {author} {\bibfnamefont {M.}~\bibnamefont {Kudra}}, \bibinfo {author} {\bibfnamefont {M.}~\bibnamefont {Kervinen}}, \bibinfo {author} {\bibfnamefont {I.}~\bibnamefont {Strandberg}}, \bibinfo {author} {\bibfnamefont {S.}~\bibnamefont {Ahmed}}, \bibinfo {author} {\bibfnamefont {M.}~\bibnamefont {Scigliuzzo}}, \bibinfo {author} {\bibfnamefont {A.}~\bibnamefont {Osman}}, \bibinfo {author} {\bibfnamefont {D.~P.}\ \bibnamefont {Lozano}}, \bibinfo {author} {\bibfnamefont {M.~O.}\ \bibnamefont {Thol\'en}}, \bibinfo {author} {\bibfnamefont {R.}~\bibnamefont {Borgani}}, \bibinfo {author} {\bibfnamefont {D.~B.}\ \bibnamefont {Haviland}}, \bibinfo {author} {\bibfnamefont {G.}~\bibnamefont {Ferrini}}, \bibinfo {author} {\bibfnamefont {J.}~\bibnamefont {Bylander}}, \bibinfo {author} {\bibfnamefont {A.~F.}\ \bibnamefont {Kockum}}, \bibinfo {author} {\bibfnamefont {F.}~\bibnamefont {Quijandr\'{\i}a}}, \bibinfo {author} {\bibfnamefont {P.}~\bibnamefont {Delsing}},\ and\ \bibinfo {author} {\bibfnamefont {S.}~\bibnamefont {Gasparinetti}},\ }\bibfield  {title} {\bibinfo {title} {{Robust Preparation of Wigner-Negative States with Optimized SNAP-Displacement Sequences}},\ }\href {https://doi.org/10.1103/PRXQuantum.3.030301} {\bibfield  {journal} {\bibinfo  {journal} {PRX Quantum}\ }\textbf {\bibinfo {volume} {3}},\ \bibinfo {pages} {030301} (\bibinfo {year} {2022})}\BibitemShut {NoStop}%
\bibitem [{\citenamefont {Cordero}\ \emph {et~al.}(2019)\citenamefont {Cordero}, \citenamefont {Nahmad-Achar}, \citenamefont {Casta\~nos},\ and\ \citenamefont {L\'opez-Pe\~na}}]{Cordero2019a}%
  \BibitemOpen
  \bibfield  {author} {\bibinfo {author} {\bibfnamefont {S.}~\bibnamefont {Cordero}}, \bibinfo {author} {\bibfnamefont {E.}~\bibnamefont {Nahmad-Achar}}, \bibinfo {author} {\bibfnamefont {O.}~\bibnamefont {Casta\~nos}},\ and\ \bibinfo {author} {\bibfnamefont {R.}~\bibnamefont {L\'opez-Pe\~na}},\ }\bibfield  {title} {\bibinfo {title} {{Optimal basis for the generalized Dicke model}},\ }\href {https://doi.org/10.1103/PhysRevA.100.053810} {\bibfield  {journal} {\bibinfo  {journal} {Phys. Rev. A}\ }\textbf {\bibinfo {volume} {100}},\ \bibinfo {pages} {053810} (\bibinfo {year} {2019})}\BibitemShut {NoStop}%
\bibitem [{\citenamefont {Rojan}\ \emph {et~al.}(2014)\citenamefont {Rojan}, \citenamefont {Reich}, \citenamefont {Dotsenko}, \citenamefont {Raimond}, \citenamefont {Koch},\ and\ \citenamefont {Morigi}}]{Rojan2014a}%
  \BibitemOpen
  \bibfield  {author} {\bibinfo {author} {\bibfnamefont {K.}~\bibnamefont {Rojan}}, \bibinfo {author} {\bibfnamefont {D.~M.}\ \bibnamefont {Reich}}, \bibinfo {author} {\bibfnamefont {I.}~\bibnamefont {Dotsenko}}, \bibinfo {author} {\bibfnamefont {J.-M.}\ \bibnamefont {Raimond}}, \bibinfo {author} {\bibfnamefont {C.~P.}\ \bibnamefont {Koch}},\ and\ \bibinfo {author} {\bibfnamefont {G.}~\bibnamefont {Morigi}},\ }\bibfield  {title} {\bibinfo {title} {{Arbitrary-quantum-state preparation of a harmonic oscillator via optimal control}},\ }\href {https://doi.org/10.1103/PhysRevA.90.023824} {\bibfield  {journal} {\bibinfo  {journal} {Phys. Rev. A}\ }\textbf {\bibinfo {volume} {90}},\ \bibinfo {pages} {023824} (\bibinfo {year} {2014})}\BibitemShut {NoStop}%
\bibitem [{\citenamefont {Yu}\ and\ \citenamefont {Luo}(2023)}]{Yu2023a}%
  \BibitemOpen
  \bibfield  {author} {\bibinfo {author} {\bibfnamefont {Z.}~\bibnamefont {Yu}}\ and\ \bibinfo {author} {\bibfnamefont {D.-W.}\ \bibnamefont {Luo}},\ }\bibfield  {title} {\bibinfo {title} {{Many-body quantum state control in the presence of environmental noise}},\ }\href {https://doi.org/10.26421/qic23.11-12-3} {\bibfield  {journal} {\bibinfo  {journal} {Quantum Information and Computation}\ }\textbf {\bibinfo {volume} {23}},\ \bibinfo {pages} {937} (\bibinfo {year} {2023})}\BibitemShut {NoStop}%
\bibitem [{\citenamefont {Riaz}\ \emph {et~al.}(2019)\citenamefont {Riaz}, \citenamefont {Shuang},\ and\ \citenamefont {Qamar}}]{Riaz2019a}%
  \BibitemOpen
  \bibfield  {author} {\bibinfo {author} {\bibfnamefont {B.}~\bibnamefont {Riaz}}, \bibinfo {author} {\bibfnamefont {C.}~\bibnamefont {Shuang}},\ and\ \bibinfo {author} {\bibfnamefont {S.}~\bibnamefont {Qamar}},\ }\bibfield  {title} {\bibinfo {title} {{Optimal control methods for quantum gate preparation: a comparative study}},\ }\href {https://doi.org/10.1007/s11128-019-2190-0} {\bibfield  {journal} {\bibinfo  {journal} {Quantum Information Processing}\ }\textbf {\bibinfo {volume} {18}},\ \bibinfo {pages} {100} (\bibinfo {year} {2019})}\BibitemShut {NoStop}%
\bibitem [{\citenamefont {Luo}\ and\ \citenamefont {Yu}(2023)}]{Luo2023a}%
  \BibitemOpen
  \bibfield  {author} {\bibinfo {author} {\bibfnamefont {D.-W.}\ \bibnamefont {Luo}}\ and\ \bibinfo {author} {\bibfnamefont {T.}~\bibnamefont {Yu}},\ }\href@noop {} {\bibinfo {title} {{Time-reversal assisted quantum metrology with an optimal control}}} (\bibinfo {year} {2023}),\ \Eprint {https://arxiv.org/abs/2312.14443} {arXiv:2312.14443 [quant-ph]} \BibitemShut {NoStop}%
\bibitem [{\citenamefont {Halaski}\ \emph {et~al.}(2024)\citenamefont {Halaski}, \citenamefont {Krauss}, \citenamefont {Basilewitsch},\ and\ \citenamefont {Koch}}]{Halaski2024a}%
  \BibitemOpen
  \bibfield  {author} {\bibinfo {author} {\bibfnamefont {A.}~\bibnamefont {Halaski}}, \bibinfo {author} {\bibfnamefont {M.~G.}\ \bibnamefont {Krauss}}, \bibinfo {author} {\bibfnamefont {D.}~\bibnamefont {Basilewitsch}},\ and\ \bibinfo {author} {\bibfnamefont {C.~P.}\ \bibnamefont {Koch}},\ }\bibfield  {title} {\bibinfo {title} {Quantum optimal control of squeezing in cavity optomechanics},\ }\href {https://doi.org/10.1103/PhysRevA.110.013512} {\bibfield  {journal} {\bibinfo  {journal} {Phys. Rev. A}\ }\textbf {\bibinfo {volume} {110}},\ \bibinfo {pages} {013512} (\bibinfo {year} {2024})}\BibitemShut {NoStop}%
\bibitem [{\citenamefont {Basilewitsch}\ \emph {et~al.}(2022)\citenamefont {Basilewitsch}, \citenamefont {Zhang}, \citenamefont {Girvin},\ and\ \citenamefont {Koch}}]{Basilewitsch2022a}%
  \BibitemOpen
  \bibfield  {author} {\bibinfo {author} {\bibfnamefont {D.}~\bibnamefont {Basilewitsch}}, \bibinfo {author} {\bibfnamefont {Y.}~\bibnamefont {Zhang}}, \bibinfo {author} {\bibfnamefont {S.~M.}\ \bibnamefont {Girvin}},\ and\ \bibinfo {author} {\bibfnamefont {C.~P.}\ \bibnamefont {Koch}},\ }\bibfield  {title} {\bibinfo {title} {Engineering strong beamsplitter interaction between bosonic modes via quantum optimal control theory},\ }\href {https://doi.org/10.1103/PhysRevResearch.4.023054} {\bibfield  {journal} {\bibinfo  {journal} {Phys. Rev. Res.}\ }\textbf {\bibinfo {volume} {4}},\ \bibinfo {pages} {023054} (\bibinfo {year} {2022})}\BibitemShut {NoStop}%
\bibitem [{\citenamefont {Ohtsuki}\ \emph {et~al.}(2003)\citenamefont {Ohtsuki}, \citenamefont {Nakagami}, \citenamefont {Zhu},\ and\ \citenamefont {Rabitz}}]{Ohtsuki2003a}%
  \BibitemOpen
  \bibfield  {author} {\bibinfo {author} {\bibfnamefont {Y.}~\bibnamefont {Ohtsuki}}, \bibinfo {author} {\bibfnamefont {K.}~\bibnamefont {Nakagami}}, \bibinfo {author} {\bibfnamefont {W.}~\bibnamefont {Zhu}},\ and\ \bibinfo {author} {\bibfnamefont {H.}~\bibnamefont {Rabitz}},\ }\bibfield  {title} {\bibinfo {title} {{Quantum optimal control of wave packet dynamics under the influence of dissipation}},\ }\href {https://doi.org/https://doi.org/10.1016/S0301-0104(02)00991-6} {\bibfield  {journal} {\bibinfo  {journal} {Chem. Phys.}\ }\textbf {\bibinfo {volume} {287}},\ \bibinfo {pages} {197} (\bibinfo {year} {2003})}\BibitemShut {NoStop}%
\bibitem [{\citenamefont {Pfister}\ \emph {et~al.}(2004)\citenamefont {Pfister}, \citenamefont {Feng}, \citenamefont {Jennings}, \citenamefont {Pooser},\ and\ \citenamefont {Xie}}]{Pfister2004a}%
  \BibitemOpen
  \bibfield  {author} {\bibinfo {author} {\bibfnamefont {O.}~\bibnamefont {Pfister}}, \bibinfo {author} {\bibfnamefont {S.}~\bibnamefont {Feng}}, \bibinfo {author} {\bibfnamefont {G.}~\bibnamefont {Jennings}}, \bibinfo {author} {\bibfnamefont {R.}~\bibnamefont {Pooser}},\ and\ \bibinfo {author} {\bibfnamefont {D.}~\bibnamefont {Xie}},\ }\bibfield  {title} {\bibinfo {title} {{Multipartite continuous-variable entanglement from concurrent nonlinearities}},\ }\href {https://doi.org/10.1103/PhysRevA.70.020302} {\bibfield  {journal} {\bibinfo  {journal} {Phys. Rev. A}\ }\textbf {\bibinfo {volume} {70}},\ \bibinfo {pages} {020302} (\bibinfo {year} {2004})}\BibitemShut {NoStop}%
\bibitem [{\citenamefont {Adesso}\ and\ \citenamefont {Illuminati}(2007)}]{Adesso2007a}%
  \BibitemOpen
  \bibfield  {author} {\bibinfo {author} {\bibfnamefont {G.}~\bibnamefont {Adesso}}\ and\ \bibinfo {author} {\bibfnamefont {F.}~\bibnamefont {Illuminati}},\ }\bibfield  {title} {\bibinfo {title} {{Entanglement in continuous-variable systems: recent advances and current perspectives}},\ }\href {https://doi.org/10.1088/1751-8113/40/28/s01} {\bibfield  {journal} {\bibinfo  {journal} {J. Phys. A}\ }\textbf {\bibinfo {volume} {40}},\ \bibinfo {pages} {7821} (\bibinfo {year} {2007})}\BibitemShut {NoStop}%
\bibitem [{\citenamefont {He}\ \emph {et~al.}(2012)\citenamefont {He}, \citenamefont {Drummond}, \citenamefont {Olsen},\ and\ \citenamefont {Reid}}]{He2012a}%
  \BibitemOpen
  \bibfield  {author} {\bibinfo {author} {\bibfnamefont {Q.~Y.}\ \bibnamefont {He}}, \bibinfo {author} {\bibfnamefont {P.~D.}\ \bibnamefont {Drummond}}, \bibinfo {author} {\bibfnamefont {M.~K.}\ \bibnamefont {Olsen}},\ and\ \bibinfo {author} {\bibfnamefont {M.~D.}\ \bibnamefont {Reid}},\ }\bibfield  {title} {\bibinfo {title} {{Einstein-Podolsky-Rosen entanglement and steering in two-well Bose-Einstein-condensate ground states}},\ }\href {https://doi.org/10.1103/PhysRevA.86.023626} {\bibfield  {journal} {\bibinfo  {journal} {Phys. Rev. A}\ }\textbf {\bibinfo {volume} {86}},\ \bibinfo {pages} {023626} (\bibinfo {year} {2012})}\BibitemShut {NoStop}%
\bibitem [{\citenamefont {Plenio}\ \emph {et~al.}(2004)\citenamefont {Plenio}, \citenamefont {Hartley},\ and\ \citenamefont {Eisert}}]{Plenio2004a}%
  \BibitemOpen
  \bibfield  {author} {\bibinfo {author} {\bibfnamefont {M.~B.}\ \bibnamefont {Plenio}}, \bibinfo {author} {\bibfnamefont {J.}~\bibnamefont {Hartley}},\ and\ \bibinfo {author} {\bibfnamefont {J.}~\bibnamefont {Eisert}},\ }\bibfield  {title} {\bibinfo {title} {{Dynamics and manipulation of entanglement in coupled harmonic systems with many degrees of freedom}},\ }\href {https://doi.org/10.1088/1367-2630/6/1/036} {\bibfield  {journal} {\bibinfo  {journal} {New Journal of Physics}\ }\textbf {\bibinfo {volume} {6}},\ \bibinfo {pages} {36} (\bibinfo {year} {2004})}\BibitemShut {NoStop}%
\bibitem [{\citenamefont {Law}(1994)}]{law1994effective}%
  \BibitemOpen
  \bibfield  {author} {\bibinfo {author} {\bibfnamefont {C.}~\bibnamefont {Law}},\ }\bibfield  {title} {\bibinfo {title} {{Effective Hamiltonian for the radiation in a cavity with a moving mirror and a time-varying dielectric medium}},\ }\href@noop {} {\bibfield  {journal} {\bibinfo  {journal} {Phys. Rev. A}\ }\textbf {\bibinfo {volume} {49}},\ \bibinfo {pages} {433} (\bibinfo {year} {1994})}\BibitemShut {NoStop}%
\bibitem [{\citenamefont {Milne}\ \emph {et~al.}(2020)\citenamefont {Milne}, \citenamefont {Edmunds}, \citenamefont {Hempel}, \citenamefont {Roy}, \citenamefont {Mavadia},\ and\ \citenamefont {Biercuk}}]{Milne2020a}%
  \BibitemOpen
  \bibfield  {author} {\bibinfo {author} {\bibfnamefont {A.~R.}\ \bibnamefont {Milne}}, \bibinfo {author} {\bibfnamefont {C.~L.}\ \bibnamefont {Edmunds}}, \bibinfo {author} {\bibfnamefont {C.}~\bibnamefont {Hempel}}, \bibinfo {author} {\bibfnamefont {F.}~\bibnamefont {Roy}}, \bibinfo {author} {\bibfnamefont {S.}~\bibnamefont {Mavadia}},\ and\ \bibinfo {author} {\bibfnamefont {M.~J.}\ \bibnamefont {Biercuk}},\ }\bibfield  {title} {\bibinfo {title} {{Phase-Modulated Entangling Gates Robust to Static and Time-Varying Errors}},\ }\href {https://doi.org/10.1103/PhysRevApplied.13.024022} {\bibfield  {journal} {\bibinfo  {journal} {Phys. Rev. Appl.}\ }\textbf {\bibinfo {volume} {13}},\ \bibinfo {pages} {024022} (\bibinfo {year} {2020})}\BibitemShut {NoStop}%
\bibitem [{\citenamefont {Bishof}\ \emph {et~al.}(2013)\citenamefont {Bishof}, \citenamefont {Zhang}, \citenamefont {Martin},\ and\ \citenamefont {Ye}}]{Bishof2013a}%
  \BibitemOpen
  \bibfield  {author} {\bibinfo {author} {\bibfnamefont {M.}~\bibnamefont {Bishof}}, \bibinfo {author} {\bibfnamefont {X.}~\bibnamefont {Zhang}}, \bibinfo {author} {\bibfnamefont {M.~J.}\ \bibnamefont {Martin}},\ and\ \bibinfo {author} {\bibfnamefont {J.}~\bibnamefont {Ye}},\ }\bibfield  {title} {\bibinfo {title} {{Optical Spectrum Analyzer with Quantum-Limited Noise Floor}},\ }\href {https://doi.org/10.1103/PhysRevLett.111.093604} {\bibfield  {journal} {\bibinfo  {journal} {Phys. Rev. Lett.}\ }\textbf {\bibinfo {volume} {111}},\ \bibinfo {pages} {093604} (\bibinfo {year} {2013})}\BibitemShut {NoStop}%
\bibitem [{\citenamefont {Ekert}\ and\ \citenamefont {Knight}(1989)}]{Ekert1989a}%
  \BibitemOpen
  \bibfield  {author} {\bibinfo {author} {\bibfnamefont {A.~K.}\ \bibnamefont {Ekert}}\ and\ \bibinfo {author} {\bibfnamefont {P.~L.}\ \bibnamefont {Knight}},\ }\bibfield  {title} {\bibinfo {title} {{Correlations and squeezing of two-mode oscillations}},\ }\href {https://doi.org/10.1119/1.15922} {\bibfield  {journal} {\bibinfo  {journal} {Am. J. Phys.}\ }\textbf {\bibinfo {volume} {57}},\ \bibinfo {pages} {692} (\bibinfo {year} {1989})}\BibitemShut {NoStop}%
\bibitem [{\citenamefont {Vidal}\ and\ \citenamefont {Werner}(2002)}]{Vidal2002a}%
  \BibitemOpen
  \bibfield  {author} {\bibinfo {author} {\bibfnamefont {G.}~\bibnamefont {Vidal}}\ and\ \bibinfo {author} {\bibfnamefont {R.~F.}\ \bibnamefont {Werner}},\ }\bibfield  {title} {\bibinfo {title} {{Computable measure of entanglement}},\ }\href {https://doi.org/10.1103/PhysRevA.65.032314} {\bibfield  {journal} {\bibinfo  {journal} {Phys. Rev. A}\ }\textbf {\bibinfo {volume} {65}},\ \bibinfo {pages} {032314} (\bibinfo {year} {2002})}\BibitemShut {NoStop}%
\bibitem [{\citenamefont {Horodecki}\ \emph {et~al.}(2009)\citenamefont {Horodecki}, \citenamefont {Horodecki}, \citenamefont {Horodecki},\ and\ \citenamefont {Horodecki}}]{Horodecki2009a}%
  \BibitemOpen
  \bibfield  {author} {\bibinfo {author} {\bibfnamefont {R.}~\bibnamefont {Horodecki}}, \bibinfo {author} {\bibfnamefont {P.}~\bibnamefont {Horodecki}}, \bibinfo {author} {\bibfnamefont {M.}~\bibnamefont {Horodecki}},\ and\ \bibinfo {author} {\bibfnamefont {K.}~\bibnamefont {Horodecki}},\ }\bibfield  {title} {\bibinfo {title} {{Quantum entanglement}},\ }\href {https://doi.org/10.1103/RevModPhys.81.865} {\bibfield  {journal} {\bibinfo  {journal} {Rev. Mod. Phys.}\ }\textbf {\bibinfo {volume} {81}},\ \bibinfo {pages} {865} (\bibinfo {year} {2009})}\BibitemShut {NoStop}%
\bibitem [{\citenamefont {Simon}(2000)}]{Simon2000a}%
  \BibitemOpen
  \bibfield  {author} {\bibinfo {author} {\bibfnamefont {R.}~\bibnamefont {Simon}},\ }\bibfield  {title} {\bibinfo {title} {{Peres-Horodecki Separability Criterion for Continuous Variable Systems}},\ }\href {https://doi.org/10.1103/PhysRevLett.84.2726} {\bibfield  {journal} {\bibinfo  {journal} {Phys. Rev. Lett.}\ }\textbf {\bibinfo {volume} {84}},\ \bibinfo {pages} {2726} (\bibinfo {year} {2000})}\BibitemShut {NoStop}%
\bibitem [{\citenamefont {Plenio}(2005)}]{Plenio2005a}%
  \BibitemOpen
  \bibfield  {author} {\bibinfo {author} {\bibfnamefont {M.~B.}\ \bibnamefont {Plenio}},\ }\bibfield  {title} {\bibinfo {title} {{Logarithmic Negativity: A Full Entanglement Monotone That is not Convex}},\ }\href {https://doi.org/10.1103/PhysRevLett.95.090503} {\bibfield  {journal} {\bibinfo  {journal} {Phys. Rev. Lett.}\ }\textbf {\bibinfo {volume} {95}},\ \bibinfo {pages} {090503} (\bibinfo {year} {2005})}\BibitemShut {NoStop}%
\bibitem [{\citenamefont {Deffner}\ and\ \citenamefont {Campbell}(2017)}]{Deffner2017a}%
  \BibitemOpen
  \bibfield  {author} {\bibinfo {author} {\bibfnamefont {S.}~\bibnamefont {Deffner}}\ and\ \bibinfo {author} {\bibfnamefont {S.}~\bibnamefont {Campbell}},\ }\bibfield  {title} {\bibinfo {title} {Quantum speed limits: from heisenberg's uncertainty principle to optimal quantum control},\ }\href {https://doi.org/10.1088/1751-8121/aa86c6} {\bibfield  {journal} {\bibinfo  {journal} {Journal of Physics A: Mathematical and Theoretical}\ }\textbf {\bibinfo {volume} {50}},\ \bibinfo {pages} {453001} (\bibinfo {year} {2017})}\BibitemShut {NoStop}%
\bibitem [{\citenamefont {Mandelstam}\ and\ \citenamefont {Tamm}(1991)}]{Mandelstam1991a}%
  \BibitemOpen
  \bibfield  {author} {\bibinfo {author} {\bibfnamefont {L.}~\bibnamefont {Mandelstam}}\ and\ \bibinfo {author} {\bibfnamefont {I.}~\bibnamefont {Tamm}},\ }\bibinfo {title} {The uncertainty relation between energy and time in non-relativistic quantum mechanics},\ in\ \href {https://doi.org/10.1007/978-3-642-74626-0_8} {\emph {\bibinfo {booktitle} {Selected Papers}}},\ \bibinfo {editor} {edited by\ \bibinfo {editor} {\bibfnamefont {B.~M.}\ \bibnamefont {Bolotovskii}}, \bibinfo {editor} {\bibfnamefont {V.~Y.}\ \bibnamefont {Frenkel}},\ and\ \bibinfo {editor} {\bibfnamefont {R.}~\bibnamefont {Peierls}}}\ (\bibinfo  {publisher} {Springer Berlin Heidelberg},\ \bibinfo {address} {Berlin, Heidelberg},\ \bibinfo {year} {1991})\ pp.\ \bibinfo {pages} {115--123}\BibitemShut {NoStop}%
\bibitem [{\citenamefont {Bhattacharyya}(1983)}]{Bhattacharyya1983a}%
  \BibitemOpen
  \bibfield  {author} {\bibinfo {author} {\bibfnamefont {K.}~\bibnamefont {Bhattacharyya}},\ }\bibfield  {title} {\bibinfo {title} {Quantum decay and the mandelstam-tamm-energy inequality},\ }\href {https://doi.org/10.1088/0305-4470/16/13/021} {\bibfield  {journal} {\bibinfo  {journal} {J. Phys. A}\ }\textbf {\bibinfo {volume} {16}},\ \bibinfo {pages} {2993} (\bibinfo {year} {1983})}\BibitemShut {NoStop}%
\bibitem [{\citenamefont {Taddei}\ \emph {et~al.}(2013)\citenamefont {Taddei}, \citenamefont {Escher}, \citenamefont {Davidovich},\ and\ \citenamefont {de~Matos~Filho}}]{Taddei2013a}%
  \BibitemOpen
  \bibfield  {author} {\bibinfo {author} {\bibfnamefont {M.~M.}\ \bibnamefont {Taddei}}, \bibinfo {author} {\bibfnamefont {B.~M.}\ \bibnamefont {Escher}}, \bibinfo {author} {\bibfnamefont {L.}~\bibnamefont {Davidovich}},\ and\ \bibinfo {author} {\bibfnamefont {R.~L.}\ \bibnamefont {de~Matos~Filho}},\ }\bibfield  {title} {\bibinfo {title} {Quantum speed limit for physical processes},\ }\href {https://doi.org/10.1103/PhysRevLett.110.050402} {\bibfield  {journal} {\bibinfo  {journal} {Phys. Rev. Lett.}\ }\textbf {\bibinfo {volume} {110}},\ \bibinfo {pages} {050402} (\bibinfo {year} {2013})}\BibitemShut {NoStop}%
\bibitem [{\citenamefont {Gajdacz}\ \emph {et~al.}(2015)\citenamefont {Gajdacz}, \citenamefont {Das}, \citenamefont {Arlt}, \citenamefont {Sherson},\ and\ \citenamefont {Opatrn\'y}}]{Gajdacz2015a}%
  \BibitemOpen
  \bibfield  {author} {\bibinfo {author} {\bibfnamefont {M.}~\bibnamefont {Gajdacz}}, \bibinfo {author} {\bibfnamefont {K.~K.}\ \bibnamefont {Das}}, \bibinfo {author} {\bibfnamefont {J.}~\bibnamefont {Arlt}}, \bibinfo {author} {\bibfnamefont {J.~F.}\ \bibnamefont {Sherson}},\ and\ \bibinfo {author} {\bibfnamefont {T.~c.~v.}\ \bibnamefont {Opatrn\'y}},\ }\bibfield  {title} {\bibinfo {title} {Time-limited optimal dynamics beyond the quantum speed limit},\ }\href {https://doi.org/10.1103/PhysRevA.92.062106} {\bibfield  {journal} {\bibinfo  {journal} {Phys. Rev. A}\ }\textbf {\bibinfo {volume} {92}},\ \bibinfo {pages} {062106} (\bibinfo {year} {2015})}\BibitemShut {NoStop}%
\bibitem [{\citenamefont {Caneva}\ \emph {et~al.}(2009)\citenamefont {Caneva}, \citenamefont {Murphy}, \citenamefont {Calarco}, \citenamefont {Fazio}, \citenamefont {Montangero}, \citenamefont {Giovannetti},\ and\ \citenamefont {Santoro}}]{Caneva2009a}%
  \BibitemOpen
  \bibfield  {author} {\bibinfo {author} {\bibfnamefont {T.}~\bibnamefont {Caneva}}, \bibinfo {author} {\bibfnamefont {M.}~\bibnamefont {Murphy}}, \bibinfo {author} {\bibfnamefont {T.}~\bibnamefont {Calarco}}, \bibinfo {author} {\bibfnamefont {R.}~\bibnamefont {Fazio}}, \bibinfo {author} {\bibfnamefont {S.}~\bibnamefont {Montangero}}, \bibinfo {author} {\bibfnamefont {V.}~\bibnamefont {Giovannetti}},\ and\ \bibinfo {author} {\bibfnamefont {G.~E.}\ \bibnamefont {Santoro}},\ }\bibfield  {title} {\bibinfo {title} {Optimal control at the quantum speed limit},\ }\href {https://doi.org/10.1103/PhysRevLett.103.240501} {\bibfield  {journal} {\bibinfo  {journal} {Phys. Rev. Lett.}\ }\textbf {\bibinfo {volume} {103}},\ \bibinfo {pages} {240501} (\bibinfo {year} {2009})}\BibitemShut {NoStop}%
\bibitem [{\citenamefont {Schumaker}\ and\ \citenamefont {Caves}(1985)}]{Schumaker1985a}%
  \BibitemOpen
  \bibfield  {author} {\bibinfo {author} {\bibfnamefont {B.~L.}\ \bibnamefont {Schumaker}}\ and\ \bibinfo {author} {\bibfnamefont {C.~M.}\ \bibnamefont {Caves}},\ }\bibfield  {title} {\bibinfo {title} {New formalism for two-photon quantum optics. ii. mathematical foundation and compact notation},\ }\href {https://doi.org/10.1103/PhysRevA.31.3093} {\bibfield  {journal} {\bibinfo  {journal} {Phys. Rev. A}\ }\textbf {\bibinfo {volume} {31}},\ \bibinfo {pages} {3093} (\bibinfo {year} {1985})}\BibitemShut {NoStop}%
\bibitem [{\citenamefont {Reich}\ \emph {et~al.}(2013)\citenamefont {Reich}, \citenamefont {Palao},\ and\ \citenamefont {Koch}}]{Reich2013a}%
  \BibitemOpen
  \bibfield  {author} {\bibinfo {author} {\bibfnamefont {D.~M.}\ \bibnamefont {Reich}}, \bibinfo {author} {\bibfnamefont {J.~P.}\ \bibnamefont {Palao}},\ and\ \bibinfo {author} {\bibfnamefont {C.~P.}\ \bibnamefont {Koch}},\ }\bibfield  {title} {\bibinfo {title} {{Optimal control under spectral constraints: enforcing multi-photon absorption pathways}},\ }\href {https://doi.org/10.1080/09500340.2013.844866} {\bibfield  {journal} {\bibinfo  {journal} {Journal of Modern Optics}\ }\textbf {\bibinfo {volume} {61}},\ \bibinfo {pages} {822} (\bibinfo {year} {2013})}\BibitemShut {NoStop}%
\bibitem [{\citenamefont {Yu}\ and\ \citenamefont {Eberly}(2009)}]{Yu2009a}%
  \BibitemOpen
  \bibfield  {author} {\bibinfo {author} {\bibfnamefont {T.}~\bibnamefont {Yu}}\ and\ \bibinfo {author} {\bibfnamefont {J.~H.}\ \bibnamefont {Eberly}},\ }\bibfield  {title} {\bibinfo {title} {{Sudden Death of Entanglement}},\ }\href {https://doi.org/10.1126/science.1167343} {\bibfield  {journal} {\bibinfo  {journal} {Science}\ }\textbf {\bibinfo {volume} {323}},\ \bibinfo {pages} {598} (\bibinfo {year} {2009})}\BibitemShut {NoStop}%
\bibitem [{\citenamefont {Yu}\ and\ \citenamefont {Eberly}(2006)}]{Yu2006a}%
  \BibitemOpen
  \bibfield  {author} {\bibinfo {author} {\bibfnamefont {T.}~\bibnamefont {Yu}}\ and\ \bibinfo {author} {\bibfnamefont {J.}~\bibnamefont {Eberly}},\ }\bibfield  {title} {\bibinfo {title} {{Sudden death of entanglement: classical noise effects}},\ }\href@noop {} {\bibfield  {journal} {\bibinfo  {journal} {Optics Communications}\ }\textbf {\bibinfo {volume} {264}},\ \bibinfo {pages} {393} (\bibinfo {year} {2006})}\BibitemShut {NoStop}%
\bibitem [{\citenamefont {Di{\'o}si}\ \emph {et~al.}(1998)\citenamefont {Di{\'o}si}, \citenamefont {Gisin},\ and\ \citenamefont {Strunz}}]{diosi1998non}%
  \BibitemOpen
  \bibfield  {author} {\bibinfo {author} {\bibfnamefont {L.}~\bibnamefont {Di{\'o}si}}, \bibinfo {author} {\bibfnamefont {N.}~\bibnamefont {Gisin}},\ and\ \bibinfo {author} {\bibfnamefont {W.~T.}\ \bibnamefont {Strunz}},\ }\bibfield  {title} {\bibinfo {title} {{Non-Markovian quantum state diffusion}},\ }\href@noop {} {\bibfield  {journal} {\bibinfo  {journal} {Phys. Rev. A}\ }\textbf {\bibinfo {volume} {58}},\ \bibinfo {pages} {1699} (\bibinfo {year} {1998})}\BibitemShut {NoStop}%
\bibitem [{\citenamefont {Strunz}\ \emph {et~al.}(1999)\citenamefont {Strunz}, \citenamefont {Diosi},\ and\ \citenamefont {Gisin}}]{strunz1999open}%
  \BibitemOpen
  \bibfield  {author} {\bibinfo {author} {\bibfnamefont {W.~T.}\ \bibnamefont {Strunz}}, \bibinfo {author} {\bibfnamefont {L.}~\bibnamefont {Diosi}},\ and\ \bibinfo {author} {\bibfnamefont {N.}~\bibnamefont {Gisin}},\ }\bibfield  {title} {\bibinfo {title} {{Open system dynamics with non-Markovian quantum trajectories}},\ }\href@noop {} {\bibfield  {journal} {\bibinfo  {journal} {Phys. Rev. Lett.}\ }\textbf {\bibinfo {volume} {82}},\ \bibinfo {pages} {1801} (\bibinfo {year} {1999})}\BibitemShut {NoStop}%
\bibitem [{\citenamefont {Yu}\ \emph {et~al.}(1999)\citenamefont {Yu}, \citenamefont {Di{\'o}si}, \citenamefont {Gisin},\ and\ \citenamefont {Strunz}}]{Yu1999a}%
  \BibitemOpen
  \bibfield  {author} {\bibinfo {author} {\bibfnamefont {T.}~\bibnamefont {Yu}}, \bibinfo {author} {\bibfnamefont {L.}~\bibnamefont {Di{\'o}si}}, \bibinfo {author} {\bibfnamefont {N.}~\bibnamefont {Gisin}},\ and\ \bibinfo {author} {\bibfnamefont {W.~T.}\ \bibnamefont {Strunz}},\ }\bibfield  {title} {\bibinfo {title} {{Non-Markovian quantum-state diffusion: Perturbation approach}},\ }\href@noop {} {\bibfield  {journal} {\bibinfo  {journal} {Phys. Rev. A}\ }\textbf {\bibinfo {volume} {60}},\ \bibinfo {pages} {91} (\bibinfo {year} {1999})}\BibitemShut {NoStop}%
\bibitem [{\citenamefont {Wiseman}\ and\ \citenamefont {Doherty}(2005)}]{Wiseman2005a}%
  \BibitemOpen
  \bibfield  {author} {\bibinfo {author} {\bibfnamefont {H.~M.}\ \bibnamefont {Wiseman}}\ and\ \bibinfo {author} {\bibfnamefont {A.~C.}\ \bibnamefont {Doherty}},\ }\bibfield  {title} {\bibinfo {title} {{Optimal Unravellings for Feedback Control in Linear Quantum Systems}},\ }\href {https://doi.org/10.1103/PhysRevLett.94.070405} {\bibfield  {journal} {\bibinfo  {journal} {Phys. Rev. Lett.}\ }\textbf {\bibinfo {volume} {94}},\ \bibinfo {pages} {070405} (\bibinfo {year} {2005})}\BibitemShut {NoStop}%
\bibitem [{\citenamefont {Zhang}\ and\ \citenamefont {M\o{}lmer}(2017)}]{Zhang2017a}%
  \BibitemOpen
  \bibfield  {author} {\bibinfo {author} {\bibfnamefont {J.}~\bibnamefont {Zhang}}\ and\ \bibinfo {author} {\bibfnamefont {K.}~\bibnamefont {M\o{}lmer}},\ }\bibfield  {title} {\bibinfo {title} {{Prediction and retrodiction with continuously monitored Gaussian states}},\ }\href {https://doi.org/10.1103/PhysRevA.96.062131} {\bibfield  {journal} {\bibinfo  {journal} {Phys. Rev. A}\ }\textbf {\bibinfo {volume} {96}},\ \bibinfo {pages} {062131} (\bibinfo {year} {2017})}\BibitemShut {NoStop}%
\bibitem [{\citenamefont {Dare}\ \emph {et~al.}(2024)\citenamefont {Dare}, \citenamefont {Hansen}, \citenamefont {Coroli}, \citenamefont {Johnson}, \citenamefont {Aspelmeyer},\ and\ \citenamefont {Deli\ifmmode~\acute{c}\else \'{c}\fi{}}}]{Dare2024a}%
  \BibitemOpen
  \bibfield  {author} {\bibinfo {author} {\bibfnamefont {K.}~\bibnamefont {Dare}}, \bibinfo {author} {\bibfnamefont {J.~J.}\ \bibnamefont {Hansen}}, \bibinfo {author} {\bibfnamefont {I.}~\bibnamefont {Coroli}}, \bibinfo {author} {\bibfnamefont {A.}~\bibnamefont {Johnson}}, \bibinfo {author} {\bibfnamefont {M.}~\bibnamefont {Aspelmeyer}},\ and\ \bibinfo {author} {\bibfnamefont {U.~c.~v.}\ \bibnamefont {Deli\ifmmode~\acute{c}\else \'{c}\fi{}}},\ }\bibfield  {title} {\bibinfo {title} {Ultrastrong linear optomechanical interaction},\ }\href {https://doi.org/10.1103/PhysRevResearch.6.L042025} {\bibfield  {journal} {\bibinfo  {journal} {Phys. Rev. Res.}\ }\textbf {\bibinfo {volume} {6}},\ \bibinfo {pages} {L042025} (\bibinfo {year} {2024})}\BibitemShut {NoStop}%
\bibitem [{\citenamefont {Gr{\"o}blacher}\ \emph {et~al.}(2009)\citenamefont {Gr{\"o}blacher}, \citenamefont {Hammerer}, \citenamefont {Vanner},\ and\ \citenamefont {Aspelmeyer}}]{Groblacher2009a}%
  \BibitemOpen
  \bibfield  {author} {\bibinfo {author} {\bibfnamefont {S.}~\bibnamefont {Gr{\"o}blacher}}, \bibinfo {author} {\bibfnamefont {K.}~\bibnamefont {Hammerer}}, \bibinfo {author} {\bibfnamefont {M.~R.}\ \bibnamefont {Vanner}},\ and\ \bibinfo {author} {\bibfnamefont {M.}~\bibnamefont {Aspelmeyer}},\ }\bibfield  {title} {\bibinfo {title} {Observation of strong coupling between a micromechanical resonator and an optical cavity field},\ }\href {https://doi.org/10.1038/nature08171} {\bibfield  {journal} {\bibinfo  {journal} {Nature}\ }\textbf {\bibinfo {volume} {460}},\ \bibinfo {pages} {724} (\bibinfo {year} {2009})}\BibitemShut {NoStop}%
\bibitem [{\citenamefont {de~los R{\'\i}os~Sommer}\ \emph {et~al.}(2021)\citenamefont {de~los R{\'\i}os~Sommer}, \citenamefont {Meyer},\ and\ \citenamefont {Quidant}}]{Rios-Sommer2021a}%
  \BibitemOpen
  \bibfield  {author} {\bibinfo {author} {\bibfnamefont {A.}~\bibnamefont {de~los R{\'\i}os~Sommer}}, \bibinfo {author} {\bibfnamefont {N.}~\bibnamefont {Meyer}},\ and\ \bibinfo {author} {\bibfnamefont {R.}~\bibnamefont {Quidant}},\ }\bibfield  {title} {\bibinfo {title} {Strong optomechanical coupling at room temperature by coherent scattering},\ }\href {https://doi.org/10.1038/s41467-020-20419-2} {\bibfield  {journal} {\bibinfo  {journal} {Nature Communications}\ }\textbf {\bibinfo {volume} {12}},\ \bibinfo {pages} {276} (\bibinfo {year} {2021})}\BibitemShut {NoStop}%
\bibitem [{\citenamefont {Rodriguez}\ \emph {et~al.}(2024)\citenamefont {Rodriguez}, \citenamefont {Ahmadi}, \citenamefont {Suarez}, \citenamefont {Mazurek}, \citenamefont {Barzanjeh},\ and\ \citenamefont {Horodecki}}]{Rodriguez2024a}%
  \BibitemOpen
  \bibfield  {author} {\bibinfo {author} {\bibfnamefont {R.~R.}\ \bibnamefont {Rodriguez}}, \bibinfo {author} {\bibfnamefont {B.}~\bibnamefont {Ahmadi}}, \bibinfo {author} {\bibfnamefont {G.}~\bibnamefont {Suarez}}, \bibinfo {author} {\bibfnamefont {P.}~\bibnamefont {Mazurek}}, \bibinfo {author} {\bibfnamefont {S.}~\bibnamefont {Barzanjeh}},\ and\ \bibinfo {author} {\bibfnamefont {P.}~\bibnamefont {Horodecki}},\ }\bibfield  {title} {\bibinfo {title} {Optimal quantum control of charging quantum batteries},\ }\href {https://doi.org/10.1088/1367-2630/ad3843} {\bibfield  {journal} {\bibinfo  {journal} {New Journal of Physics}\ }\textbf {\bibinfo {volume} {26}},\ \bibinfo {pages} {043004} (\bibinfo {year} {2024})}\BibitemShut {NoStop}%
\end{thebibliography}%


%


\end{document}